\documentclass[prd, twocolumn, superscriptaddress]{revtex4-2}
\usepackage{amsmath}
\usepackage{amssymb}

\usepackage{bm}

\usepackage{tikz}

\usepackage{array}

\allowdisplaybreaks[1]

\begin{document}
\title{Frame-dependence of the non-relativistic limit of quantum fields}

\author{Riccardo Falcone}
\affiliation{Department of Physics, University of Sapienza, Piazzale Aldo Moro 5, 00185 Rome, Italy}

\author{Claudio Conti}
\affiliation{Department of Physics, University of Sapienza, Piazzale Aldo Moro 5, 00185 Rome, Italy}
\affiliation{Institute for Complex Systems (ISC-CNR), Department of Physics, University Sapienza, Piazzale Aldo Moro 2, 00185, Rome, Italy}
\affiliation{Research Center Enrico Fermi, Via Panisperna 89a, 00184 Rome, Italy}

\begin{abstract}
We study the non-relativistic limit of quantum fields for an inertial and a non-inertial observer. We show that non-relativistic particle states appear as a superposition of relativistic and non-relativistic particles in different frames. Hence, the non-relativistic limit is frame-dependent. We detail this result when the non-inertial observer has uniform constant acceleration. Only for low accelerations, the accelerated observer agrees with the inertial frame about the non-relativistic nature of particles locally. In such a quasi-inertial regime, both observers agree about the number of particles describing quantum field states. The same does not occur when the acceleration is arbitrarily large (e.g., the Unruh effect). We furthermore prove that wave functions of particles in the inertial and the quasi-inertial frame are identical up to the coordinate transformation relating the two frames.
\end{abstract}

\maketitle

\section{Introduction}

Since the theoretical proposal of the Unruh effect \cite{PhysRevD.7.2850, Davies:1974th, PhysRevD.14.870} as the equivalent of the Hawking effect \cite{Hawking:1975vcx} in accelerated frames, there has been a wide interest in detectors able to reveal such effect. In their pioneering works, Unruh and DeWitt \cite{PhysRevD.14.870, PhysRevD.29.1047, hawking1980general} considered a particle in a box detecting field excitation in the comoving frame via monopole coupling. These works provided a toy model for the description of non-inertial detectors that interact with fields in their proper frame and produce acceleration-induced effects. The same model has been used in the context of electrodynamics for light-matter interaction of accelerated atoms (see e.g., \cite{PhysRevLett.91.243004, PhysRevA.74.023807, PhysRevLett.128.163603}).

Atomic Unruh-DeWitt detectors are described by a first-quantization prescription: the atom is assumed to be non-relativistic and made by a fixed number of particles. However, to the best of our knowledge, the fact that such description is frame-dependent has been overlooked. Remarkably, one has to take into account that the laboratory and the atom frame have different representations for the same quantum system.

One of the features of quantum field theory in curved spacetime is the fact that different observers give different particle representations for the same state \cite{wald1994quantum}. For instance, the vacuum state of an inertial frame appears as a thermal bath of particles if seen by accelerated observers \cite{PhysRevD.7.2850, Davies:1974th, PhysRevD.14.870}. As a consequence of such frame-dependence, the number of particles is generally not preserved if one switches from one frame to another. This poses a problem for the first quantization description of atomic systems in non-inertial frames. An accelerating atom --- that is prepared in the laboratory frame with a fixed number of electrons --- appears as made by an indefinite number of particles in its proper frame.

In addition to fixed numbers of particles, non-relativistic energies are assumed for the first-quantization of atomic systems. One may wonder if, along with the number of particles, the non-relativistic limit is a frame-dependent feature of the quantum system. To address such question, in this manuscript we investigate the non-relativistic limit of fields in different frames.

In our previous work \cite{falcone2022non}, we derived the non-relativistic limit of scalar and Dirac fields in curved spacetimes. Here, we study the points of view of an inertial and a non-inertial observer. We show that the two observers do not always agree about the non-relativistic nature of particles. Specifically, states that are non-relativistic in the inertial frame appear as a mixture of relativistic and non-relativistic particles if seen by the non-inertial observer. We detail the case in which the non-inertial observer is uniformly accelerated and we quantify the presence of non-relativistic particles depending on the magnitude of the acceleration $\alpha$.

As a consequence of such frame-dependence, accelerated observers cannot rely on the non-relativistic description of atomic systems. Conversely, no problem arises when both observers are inertial and moving with low relative velocities. We are, hence, motivated to look for a trade-off between non-inertial ($\alpha \neq 0$) and inertial ($\alpha = 0$) motion, which, respectively, produces acceleration-induced effects and preserves non-relativistic energies and number of particles. We show that such quasi-inertial condition is met when $\alpha$ is sufficiently small and when both the state and the non-inertial observer are localized where the metric is almost flat. In such quasi-inertial regime, the two observers agree about the non-relativistic nature and the number of particles.

In addition, we show that wave functions describing states in the quasi-inertial frame are approximated by the corresponding wave functions in the inertial frame, with the only difference coming from the coordinate transformation relating the two frames. In other words, particle states appear identical --- i.e., with the same number of particles and the same wave function --- if seen by either observers.

We detail the results by considering Gaussian single-particles and the related quasi-inertial regime. The accelerated observer sees a non-relativistic particle only when $\alpha$ is sufficiently small and the wave packet in the inertial frame is narrower than the scale length of the curvature, but wider than any relativistic wavelength. We also show that the wave function describing the state in the accelerated frame is approximately Gaussian.

The manuscript is organized in the following way. In Sec.~\ref{Inertial_and_non_inertial_frame} we consider an inertial and a non-inertial frame and show that such observers generally do not agree about the non-relativistic nature and the number of particles. In Sec.~\ref{Inertial_and_accelerated_frame} we consider the specific case of a constant uniform acceleration. The case of low acceleration and quasi-flat metric is discussed in Sec.~\ref{Inertial_and_quasiinertial_frame}. In such regime, the non-relativistic limit, number of particles and wave function of any state are approximately the same in the two frames. We detail these results in Sec.~\ref{Gaussian_singleparticle} for Gaussian single particles. Conclusions are drawn in Sec.~\ref{Conclusions}.

\section{Inertial and non-inertial frame}\label{Inertial_and_non_inertial_frame}

Here, we consider two sets of coordinates. With $(t,\vec{x})$ we refer to an inertial frame, characterized by the Minkowski metric
\begin{equation}
\eta_{\mu\nu} = \text{diag}(-c^2,1,1,1),
\end{equation}
where $c$ is the speed of light. We also consider a coordinate transformation $(t,\vec{x}) \mapsto (T,\vec{X})$ such that the frame $(T,\vec{X})$ is non-inertial and associated to a static metric $g_{\mu\nu}$. The condition of static spacetime guarantees the definition of particles with defined real energy \cite{falcone2022non}. Moreover, we consider a complex scalar field in the inertial ($\hat{\phi}$) and in the non-inertial ($\hat{\Phi}$) frame. $\hat{\phi}(t,\vec{x})$ transforms into $\hat{\Phi}(T,\vec{X})$ as a scalar field, under the coordinate transformation $(t,\vec{x}) \mapsto (T,\vec{X})$:
\begin{equation}\label{scalar_transformation}
\hat{\Phi}(T,\vec{X}) = \hat{\phi}(t(T,\vec{X}),\vec{x}(T,\vec{X})).
\end{equation}
The aim of this section is to show that the non-relativistic limit in $(t,\vec{x})$ is generally non compatible with the non-relativistic limit in $(T,\vec{X})$.

We start by decomposing $\hat{\phi}$ with respect to Klein-Gordon modes $g(\theta)$ and $ h(\theta)$:
\begin{subequations}\label{Klein_Gordon}
\begin{align}
& \left[ c^2 \eta^{\mu\nu} \partial_\mu \partial_\nu - \left( \frac{mc^2}{\hbar} \right)^2 \right] g(\theta) = 0,
\\
& \left[ c^2 \eta^{\mu\nu} \partial_\mu \partial_\nu - \left( \frac{mc^2}{\hbar} \right)^2 \right] h(\theta) = 0,
\end{align}
\end{subequations}
where $\theta$ is a discrete and/or continuum collection of quantum numbers and $g(\theta)$ and $ h(\theta)$ have, respectively, positive and negative frequencies:
\begin{align}\label{positive_negative_frequencies}
& g(\theta,t,\vec{x}) = \tilde{g}(\theta,\vec{x}) e^{-i \omega(\theta) t}, & h(\theta,t,\vec{x}) = \tilde{h}(\theta,\vec{x}) e^{i \omega(\theta) t},
\end{align}
with $\hbar \omega(\theta)$ as the energy of the single-particle with quantum numbers $\theta$. The decomposition of the field $\hat{\phi}$ reads
\begin{equation}\label{free_field_positive_negative_frequencies}
\hat{\phi}(t,\vec{x}) = \sum_\theta \left[ g(\theta, t,\vec{x}) \hat{a}(\theta)  + h(\theta, t,\vec{x}) \hat{b}^\dagger(\theta) \right],
\end{equation}
where $ \sum_\theta$ is a generalized sum, $\hat{a}(\theta)$ is the annihilation operator for the particle with quantum numbers $\theta$ and $\hat{b}^\dagger(\theta)$ is the creation operator for the associated antiparticle. 

The modes $g(\theta)$ and $ h(\theta)$ are orthonormal with respect to the Klein-Gordon inner product:
\begin{subequations}\label{KG_scalar_product_orthonormality_g}
\begin{align}
& ( g(\theta), g(\theta') )_\text{KG} = \delta_{\theta\theta'}, \\
& ( h(\theta), h(\theta') )_\text{KG} = -\delta_{\theta\theta'}, \\
& ( g(\theta), h(\theta') )_\text{KG} = 0,
\end{align}
\end{subequations}
where
\begin{align}\label{KG_scalar_product}
( \phi, \phi' )_\text{KG} = & \frac{i}{\hbar c^2} \int_{\mathbb{R}^3} d^3x \left[ \phi^*(t,\vec{x}) \partial_0  \phi'(t,\vec{x}) \right. \nonumber \\
& \left. -  \phi'(t,\vec{x}) \partial_0 \phi^*(t,\vec{x}) \right].
\end{align}
The deltas in Eq.~(\ref{KG_scalar_product_orthonormality_g}) are generalized: they act as Kronecker deltas for discrete indexes and as Dirac deltas for continuum variables.

We also define the vacuum state $| 0_\text{M} \rangle $ with respect to $\hat{a}(\theta)$ and $\hat{b}(\theta)$:
\begin{align}\label{Minkowski_vacuum}
& \hat{a}(\theta) | 0_\text{M} \rangle = 0, & \hat{b}(\theta) | 0_\text{M} \rangle = 0.
\end{align}

Similar decomposition occurs for the field $\hat{\Phi}$:
\begin{equation}\label{Klein_Gordon_curved_Phi}
\hat{\Phi}(T,\vec{X}) = \sum_\Theta\left[ G(\Theta,T,\vec{X}) \hat{A}(\Theta)  + H(\Theta,T,\vec{X}) \hat{B}^\dagger(\Theta) \right],
\end{equation}
where $G(\Theta)$ and $H(\Theta)$ are curved Klein-Gordon modes with real frequencies with respect to the time coordinate $T$:
\begin{subequations} \label{Klein_Gordon_curved}
\begin{align}
& \left[ \frac{c^2}{\sqrt{-g}} \partial_\mu \left( \sqrt{-g} g^{\mu \nu}\partial_\nu   \right) - \left( \frac{mc^2}{\hbar}\right)^2 \right] G(\Theta) = 0, \\
& \left[ \frac{c^2}{\sqrt{-g}} \partial_\mu \left( \sqrt{-g} g^{\mu \nu}\partial_\nu   \right) - \left( \frac{mc^2}{\hbar}\right)^2 \right] H(\Theta) = 0,
\end{align}
\end{subequations}
\begin{subequations}\label{scalar_curved_modes_tilde}
\begin{align} 
& G(\Theta,T,\vec{X}) = \tilde{G}(\Theta,\vec{X}) e^{-i \Omega(\Theta) T}, \\
 & H(\Theta,T,\vec{X}) = \tilde{H}(\Theta,\vec{X}) e^{i \Omega(\Theta) T}.
\end{align}
\end{subequations}
$\hat{A}(\Theta)$ ($\hat{B}(\Theta)$) is the annihilation operator associated to the particle (antiparticle) with quantum numbers $\Theta$.

In this case, the orthonormality of $G(\Theta)$ and $H(\Theta)$ modes is with respect to the curved Klein-Gordon scalar product:
\begin{subequations}\label{KG_scalar_curved_product_orthonormality_G}
\begin{align}
& ( G(\Theta), G(\Theta'))_\text{CKG} = \delta_{\Theta\Theta'} , \\
 & ( H(\Theta), H(\Theta'))_\text{CKG}  = - \delta_{\Theta\Theta'} ,\\
& ( G(\Theta), H(\Theta'))_\text{CKG} = 0.
\end{align}
\end{subequations}
Such scalar product reads
\begin{align}\label{KG_curved_scalar_product}
 & ( \Phi, \Phi' )_\text{CKG} = -\frac{i}{\hbar c} \int_{\mathbb{R}^3} d^3X  \sqrt{-g(T,\vec{X})} g^{0 \mu}(T,\vec{X}) \nonumber \\
& \times \left[ \Phi^*(T,\vec{X}) \partial_\mu \Phi'(T,\vec{X})  -  \Phi'(T,\vec{X}) \partial_\mu \Phi^*(T,\vec{X}) \right].
\end{align}

The vacuum state $| 0_\text{NM} \rangle $ of the field $\hat{\Phi}$ reads
\begin{align}
& \hat{A}(\Theta) | 0_\text{NM} \rangle = 0, & \hat{B}(\Theta) | 0_\text{NM} \rangle = 0.
\end{align}

Creation and annihilation operators of particles and antiparticles with respect to $\hat{\phi}$ and $\hat{\Phi}$ are related by a Bogoliubov transformation:
\begin{subequations}\label{Bogoliubov_transformations}
\begin{align}
& \hat{a}(\theta) = \sum_\Theta \left[ \alpha(\theta,\Theta)\hat{A}(\Theta) + \beta(\theta,\Theta) \hat{B}^\dagger(\Theta) \right], \\
& \hat{b}(\theta) = \sum_\Theta \left[ \gamma(\theta,\Theta)\hat{B}(\Theta) + \delta(\theta,\Theta) \hat{A}^\dagger(\Theta) \right].
\end{align}
\end{subequations}
A general procedure to compute Eq.~(\ref{Bogoliubov_transformations}) is the following. One starts by isolating $\hat{a}(\theta)$ and $\hat{b}^\dagger(\theta)$ from Eq.~(\ref{free_field_positive_negative_frequencies}) by using the Klein-Gordon scalar product (\ref{KG_scalar_product}) and the orthonormality conditions (\ref{KG_scalar_product_orthonormality_g}):
\begin{align}\label{Bogoliubov_transformations_1}
& \hat{a}(\theta) = ( g(\theta), \hat{\phi})_\text{KG}, & \hat{b}^\dagger(\theta) = -( h(\theta), \hat{\phi})_\text{KG}.
\end{align}
Then, one combines Eq.~(\ref{Bogoliubov_transformations_1}) with the inverse of Eq.~(\ref{scalar_transformation}) and with Eq.~(\ref{Klein_Gordon_curved_Phi}) to obtain an equation with the form of Eq.~(\ref{Bogoliubov_transformations}).

By using Eq.~(\ref{Bogoliubov_transformations}) in Eq.~(\ref{Minkowski_vacuum}), one can also derive the relation between $| 0_\text{M} \rangle $ and $| 0_\text{NM} \rangle $. The Minkowski vacuum $| 0_\text{M} \rangle $ can be written as an element of the Fock space $\mathcal{F}_\text{NM}$ generated by the vacuum state $| 0_\text{NM} \rangle $ and the creation operators $\hat{A}^\dagger(\Theta)$ and $\hat{B}^\dagger(\Theta)$. Analogously, $| 0_\text{NM} \rangle $ can be seen as an element of the Minkowski-Fock space $\mathcal{F}_\text{M}$ \footnote{We point out that this procedure is not always guaranteed. By following the algebraic approach \cite{wald1994quantum}, one finds out that a state of $\mathcal{F}_\text{NM}$ ($\mathcal{F}_\text{M}$) is not always exactly identified with an element of $\mathcal{F}_\text{M}$ ($\mathcal{F}_\text{NM}$). This occurs because we are considering two unitarily nonequivalent quantum field theory constructions. One can only guarantee the equivalence between a state of $\mathcal{F}_\text{M}$ and a state of $\mathcal{F}_\text{NM}$ up to an arbitrarily large precision --- with respect to any finite set of mean values.}.

We define the non-relativistic limit as
\begin{equation}\label{non_relativistic_limit}
\frac{\hbar \omega}{mc^2} - 1 \lesssim \epsilon
\end{equation}
in the Minkowski spacetime and
\begin{equation}\label{non_relativistic_limit_curved}
\left| \frac{\hbar \Omega}{mc^2} - 1 \right|  \lesssim \epsilon
\end{equation}
in the non-inertial frame, where $\epsilon \ll 1$ is a parameter that is vanishing in the non-relativistic limit. A non-relativistic particle in the inertial (non-inertial) frame is defined by the quantum numbers $\theta$ ($\Theta$) such that $\omega(\theta)$ ($\Omega(\Theta)$) is of order given by Eq.~(\ref{non_relativistic_limit}) (Eq.~(\ref{non_relativistic_limit_curved})). Correspondingly, a non-relativistic Fock state in the inertial (non-inertial) frame is defined by non-relativistic particles created in the vacuum state $| 0_\text{M} \rangle $ ($| 0_\text{NM} \rangle $).

One can notice that, even if $\theta$ is non-relativistic --- i.e., $\hbar \omega(\theta) / mc^2 - 1 \lesssim \epsilon$ --- the sum of Eq.~(\ref{Bogoliubov_transformations}) runs over all values of $\Theta$, including the ones such that $\Omega(\Theta)$ is relativistic --- i.e., $|\hbar \Omega(\Theta)/(mc^2) - 1| \gg \epsilon$. This means that the Bogoliubov transformation (\ref{Bogoliubov_transformations}) mixes non-relativistic modes of one frame with relativistic modes of the other. The effect is twofold. On one hand, the ``sea'' of non-inertial particles and antiparticles populating the Minkowski vacuum $| 0_\text{M} \rangle $ in $\mathcal{F}_\text{NM}$ generally includes states with relativistic energies. On the other hand, a non-relativistic particle (antiparticle) creator $\hat{a}^\dagger(\theta)$ ($\hat{b}^\dagger(\theta)$) can be responsible for the creation and the destruction of relativistic non-inertial (anti)particles. These two facts imply that an element of $\mathcal{F}_\text{M}$ that is made of non-relativistic (anti)particles, when seen as an element of $\mathcal{F}_\text{NM}$, is generally made of relativistic Minkowski (anti)particles. The other way around is also true: not always an element of $\mathcal{F}_\text{NM}$ made by non-relativistic (anti)particles is also made by non-relativistic (anti)particles in $\mathcal{F}_\text{M}$.

Given a frame of reference $K$, non-relativistic states are defined as elements of the Fock space of $K$ made of non-relativistic particles. When seen by a different observer $K'$, such states appear as a mixture of relativistic and non-relativistic particles. The conclusion is that the non-relativistic limit is frame-dependent.

One can also deduce from Eq.~(\ref{Bogoliubov_transformations}) the non conservation of particle and antiparticle number when switching from $K$ to $K'$. An element of $\mathcal{F}_\text{M}$ with $n$ particles and $m$ antiparticles is not an element of $\mathcal{F}_\text{NM}$ with the same number of particles and antiparticles. This occurs because $| 0_\text{M} \rangle $ is not a vacuum state for $\mathcal{F}_\text{NM}$ and Minkowski particle (antiparticle) creators $\hat{a}^\dagger(\theta)$ ($\hat{b}^\dagger(\theta)$) annihilate non-Minkowski antiparticles (particles), besides creating non-inertial particles (antiparticles).

\section{Inertial and accelerated frame} \label{Inertial_and_accelerated_frame}

In the present section, the non-inertial observer is assumed to have uniform acceleration $\alpha = c^2 a$ along the $x$ axis. We hence consider Rindler frames, defined by the following coordinate transformations
\begin{subequations} \label{Rindler_coordinate_transformation}
\begin{align}
& a c t_\nu = \exp (s_\nu a X) \sinh (a c T), \\ & a x_\nu = s_\nu \exp (s_\nu aX) \cosh (a c T) \label{Rindler_coordinate_transformation_x},
\end{align}
\end{subequations}
with $\nu \in \{ \text{L}, \text{R} \}$ and where $s_\text{L} = -1$ and $s_\text{R} = 1$. We also assume that $a > 0$, so that the coordinates $(t_\text{L},\vec{x}_\text{L})$ cover the left wedge defined by $x < - c |t|$ and $(t_\text{R},\vec{x}_\text{R})$ cover the region $x > c |t|$. It can be noticed that the two coordinate transformations in Eq.~(\ref{Rindler_coordinate_transformation}) differ by a sign in front of $a$: one can switch from the left to the right wedge and the other way round by letting $a \mapsto -a$.

The metric $g_{\mu\nu}$ in the right wedge reads
\begin{equation} \label{Rindler_metric}
g_{\mu\nu}(T,\vec{X}) = \text{diag} \left( -c^2 e^{2aX}, e^{2aX}, 1, 1 \right).
\end{equation}
The left wedge metric is obtained by $a \mapsto -a$ in Eq.~(\ref{Rindler_metric}).

The scalar field in the Rindler frame $\hat{\Phi}_\nu$ is related to $ \hat{\phi}$ through Eq.~(\ref{scalar_transformation}):
\begin{equation}\label{scalar_transformation_Rindler}
\hat{\Phi}_\nu(T,\vec{X}) = \hat{\phi}(t_\nu(T,\vec{X}),\vec{x}_\nu(T,\vec{X})),
\end{equation}
where the transformations $t_\nu(T,\vec{X})$, $\vec{x}_\nu(T,\vec{X})$ are given by Eq.~(\ref{Rindler_coordinate_transformation}).

Here, we consider the decomposition of the Minkowski scalar field $\hat{\phi}$ in Klein-Gordon modes with defined momenta. For such decomposition, the quantum numbers $\theta$ are the vectorial components of momenta $\vec{k}=(k_1,k_2,k_3)$. Equation (\ref{free_field_positive_negative_frequencies}) reads
\begin{equation} \label{free_field}
\hat{\phi}(t,\vec{x}) = \int_{\mathbb{R}^3} d^3 k \left[ f(\vec{k},t,\vec{x}) \hat{a}(\vec{k}) + f^*(\vec{k},t,\vec{x}) \hat{b}^\dagger(\vec{k}) \right],
\end{equation}
with
\begin{equation}\label{free_modes}
f(\vec{k},t,\vec{x}) =  \sqrt{\frac{\hbar c^2}{(2\pi)^3 2 \omega(\vec{k})}} e^{-i\omega(\vec{k})t + i\vec{k} \cdot \vec{x}}
\end{equation}
as Klein-Gordon mode with momentum $k$ and frequency
\begin{equation} \label{dispersion_relation}
\omega(\vec{k}) = \sqrt{\left(\frac{mc^2}{\hbar}\right)^2 + (c k)^2},
\end{equation}
where $k = |\vec{k}|$.

Conversely, a decomposition of $\hat{\Phi}_\text{R} $ can be obtained by considering frequency $\Omega$ and transverse momenta components $\vec{K}_\perp = (K_2,K_3)$ as quantum numbers $\vec{\Theta} = (\Omega, \vec{K}_\perp)$ \cite{RevModPhys.80.787}:
\begin{align}
\hat{\Phi}_\text{R}(T,\vec{X}) = & \int_0^{\infty} d\Omega \int_{\mathbb{R}^2} d^2 K_\perp  \nonumber \\
& \times \left[ F(\Omega,\vec{K}_\perp,T,\vec{X}) \hat{A}_\text{R}(\Omega,\vec{K}_\perp) \right. \nonumber \\
& \left. + F^*(\Omega,\vec{K}_\perp,T,\vec{X})\hat{B}_\text{R}^\dagger(\Omega,\vec{K}_\perp) \right],
\end{align}
with
\begin{subequations}
\begin{equation}
F(\Omega,\vec{K}_\perp,T,\vec{X}) = \tilde{F}(\Omega,\vec{K}_\perp,X) e^{ i \vec{K}_\perp \cdot \vec{X}_\perp - i \Omega T }, \label{F_Rindler} 
\end{equation}
\begin{align}
& \tilde{F}(\Omega,\vec{K}_\perp,X) = \frac{1}{2 \pi^2} \sqrt{ \frac{\hbar}{a} \left| \sinh \left( \frac{\beta \Omega}{2} \right) \right| } \nonumber \\
&  K_{i \Omega / (c a)} \left( \sqrt{c^2 K_\perp^2 + \left(\frac{mc^2}{\hbar}\right)^2} \frac{e^{aX}}{c a} \right), \label{F_tilde_Rindler}
\end{align}
\begin{equation}
\beta = \frac{2 \pi}{c a},
\end{equation}
\end{subequations}
and where $\vec{X}_\perp = (Y, Z)$ are the Rindler transverse coordinates. $K_\zeta (\xi)$ appearing in Eq.~(\ref{F_tilde_Rindler}) is the modified Bessel function of the second kind.

In the left wedge, $\hat{\Phi}_\text{L} $ can be decomposed as $\hat{\Phi}_\text{R}$ with $X \mapsto -X$. Indeed, the Klein-Gordon equation in Rindler spacetime
\begin{equation}\label{Rindler_Klein_Gordon}
\left\lbrace - \partial_0^2 + c^2 \partial_1^2 +   c^2 e^{2 a X} \left[ \partial_2^2 + \partial_3^2 - \left( \frac{mc}{\hbar} \right)^2 \right] \right\rbrace F(\vec{\Theta}) = 0
\end{equation}
is invariant under the transformation $a \mapsto -a$, $X \mapsto -X$ and the orthonormality condition
\begin{subequations}\label{KG_scalar_curved_product_orthonormality_F}
\begin{align}
& ( F(\vec{\Theta}), F(\vec{\Theta}'))_\text{CKG} = \delta^3 (\vec{\Theta}-\vec{\Theta}') , \\
 & ( F^*(\vec{\Theta}), F^*(\vec{\Theta}'))_\text{CKG}  = - \delta^3 (\vec{\Theta}-\vec{\Theta}') ,\\
 & ( F(\vec{\Theta}), F^*(\vec{\Theta}'))_\text{CKG} = 0
\end{align}
\end{subequations}
also holds for the modes $F(\Omega,\vec{K}_\perp,T, -X, \vec{X}_\perp)$ in the left wedge. Therefore, by considering both wedges, the field $\hat{\Phi}_\nu (T,\vec{X})$ is
\begin{align} \label{Rindler_scalar_decomposition}
\hat{\Phi}_\nu (T,\vec{X}) = & \int_0^{\infty} d\Omega \int_{\mathbb{R}^2} d^2 K_\perp   \nonumber \\
& \times  \left[ F(\Omega,\vec{K}_\perp,T,s_\nu X, \vec{X}_\perp)\hat{A}_\nu(\Omega,\vec{K}_\perp)  \right. \nonumber \\
& \left. + F^*(\Omega,\vec{K}_\perp,T,s_\nu X, \vec{X}_\perp)\hat{B}_\nu^\dagger(\Omega,\vec{K}_\perp) \right].
\end{align}

The Bogoliubov transformations relating $\hat{a}(\vec{k})$ and $\hat{b}(\vec{k})$ with $\hat{A}_\nu(\Omega,\vec{K}_\perp)$ and $\hat{B}_\nu(\Omega,\vec{K}_\perp)$ [Eq.~(\ref{Bogoliubov_transformations})] read
\begin{subequations}\label{Rindler_Bogoliubov_transformations}
\begin{align}
\hat{a}(\vec{k}) = &   \sum_{\nu=\{\text{L},\text{R}\}}\int_0^\infty d\Theta_1 \int_{\mathbb{R}^2} d^2\vec{\Theta}_\perp    \left[ \alpha_\nu(\vec{k},\vec{\Theta})  \hat{A}_\nu(\vec{\Theta}) \right. \nonumber \\
& \left.+ \alpha_\nu(\vec{k},-\vec{\Theta}) \hat{B}^\dagger_\nu(\vec{\Theta}) \right], \\
\hat{b}(\vec{k}) = &  \sum_{\nu=\{\text{L},\text{R}\}} \int_0^\infty d\Theta_1 \int_{\mathbb{R}^2} d^2\vec{\Theta}_\perp \left[ \alpha_\nu(\vec{k},\vec{\Theta})  \hat{B}_\nu(\vec{\Theta}) \right. \nonumber \\
& \left. + \alpha_\nu(\vec{k},-\vec{\Theta}) \hat{A}^\dagger_\nu(\vec{\Theta})  \right],
\end{align}
\end{subequations}
where
\begin{align} \label{alpha}
\alpha_\nu(\vec{k},\vec{\Theta}) = & \int_{\mathbb{R}^3} d^3 x \frac{\theta(s_\nu x)}{\hbar c^2}  \left[  \frac{s_\nu \Theta_1}{a x} + \omega(\vec{k}) \right] f^*( \vec{k}, 0,  \vec{x} ) \nonumber \\
& \times \tilde{F}(\vec{\Theta},  s_\nu X_\nu(x)) e^{ i \vec{\Theta}_\perp \cdot \vec{x}_\perp},
\end{align}
$\vec{x}_\perp = (y, z)$ are Minkowski transverse coordinates, $\vec{\Theta}_\perp = (\Theta_2, \Theta_3)$ the transverse coordinates of $\vec{\Theta} = (\Theta_1, \Theta_2, \Theta_3)$ and $\theta(x)$ the Heaviside theta function. The function $X_\nu(x)$ appearing in Eq.~(\ref{alpha}) is the inverse of Eq.~(\ref{Rindler_coordinate_transformation_x}) when $t = T = 0$:
\begin{equation}\label{Minkowski_to_Rindler_coordinates}
X_\nu(x) = \frac{s_\nu}{a} \ln ( s_\nu a x ).
\end{equation}
In \ref{appendix_a} we provide an explicit proof for Eqs.~(\ref{Rindler_Bogoliubov_transformations}) and (\ref{alpha}).

We write Eq.~(\ref{Rindler_Bogoliubov_transformations}) in a more compact form in the following way
\begin{subequations}\label{Rindler_Bogoliubov_transformations_compact}
\begin{align}
& \hat{a}(\vec{k}) = \sum_{\nu=\{\text{L},\text{R}\}}   \int_{\mathbb{R}^3} d^3 \Theta \alpha_\nu(\vec{k},\vec{\Theta})  \hat{\mathcal{A}}_\nu(\vec{\Theta}), \\
& \hat{b}(\vec{k}) = \sum_{\nu=\{\text{L},\text{R}\}}   \int_{\mathbb{R}^3} d^3 \Theta \alpha_\nu(\vec{k},\vec{\Theta})  \hat{\mathcal{B}}_\nu(\vec{\Theta}),
\end{align}
\end{subequations}
where
\begin{subequations}\label{Rindler_Bogoliubov_transformations_compact_AB}
\begin{align}
& \hat{\mathcal{A}}_\nu(\vec{\Theta}) = \begin{cases}
 \hat{A}_\nu(\vec{\Theta}) & \text{if } \Theta_1 > 0 \\
 \hat{B}^\dagger_\nu(-\vec{\Theta}) & \text{if } \Theta_1 < 0 \\
\end{cases} , \\
& \hat{\mathcal{B}}_\nu(\vec{\Theta}) = \begin{cases}
 \hat{B}_\nu(\vec{\Theta}) & \text{if } \Theta_1 > 0 \\
 \hat{A}^\dagger_\nu(-\vec{\Theta}) & \text{if } \Theta_1 < 0 \\
\end{cases}.
\end{align}
\end{subequations}

The Rindler vacuum state $| 0_\text{L}, 0_\text{R} \rangle$ --- which is annihilated by $\hat{A}_\nu(\Omega,\vec{K}_\perp)$ and $\hat{B}_\nu(\Omega,\vec{K}_\perp)$ operators --- and the Minkowski vacuum state $| 0_\text{M} \rangle$ are related by the following identity \cite{RevModPhys.80.787}
\begin{equation}\label{Rindler_vacuum_to_Minkowski}
| 0_\text{M} \rangle = \hat{S} | 0_\text{L}, 0_\text{R} \rangle,
\end{equation}
with the following unitary operator
\begin{align}\label{Rindler_vacuum_to_Minkowski_unitary_operator}
\hat{S} = & \exp\left(  2 \int_0^\infty d\Omega \int_{\mathbb{R}^2} d^2\vec{K}_\perp \exp \left( - \frac{\beta \Omega}{2} \right) \right. \nonumber \\
& \times \left[ \hat{A}^\dagger_\text{L}(\Omega,\vec{K}_\perp) \hat{B}^\dagger_\text{R}(\Omega,-\vec{K}_\perp)  \right. \nonumber \\
& \left. \left.   + \hat{B}^\dagger_\text{L}(\Omega,\vec{K}_\perp) \hat{A}^\dagger_\text{R}(\Omega,-\vec{K}_\perp) \right]^\text{A} \right),
\end{align}
and where $\hat{O}^\text{A} = (\hat{O}-\hat{O}^\dagger)/2$ is the antihermitian part of any operator $\hat{O}$.

Equations (\ref{Rindler_Bogoliubov_transformations}) and (\ref{Rindler_vacuum_to_Minkowski}) give the same results of Sec. \ref{Inertial_and_non_inertial_frame}: any Minkowski-Fock state $| \phi \rangle \in \mathcal{F}_\text{M}$ made of non-relativistic (anti)particles can also be seen as an element of Rindler-Fock space $\mathcal{F}_\text{LR}$ where the Minkowski vacuum background $| 0_\text{M} \rangle$ is converted into a see of Rindler (anti)particles --- including relativistic ones [Eq.~(\ref{Rindler_vacuum_to_Minkowski})] --- and any $\hat{a}^\dagger(\vec{k})$ and $\hat{b}^\dagger(\vec{k})$ operator acting on $| 0_\text{M} \rangle$ is converted into creation and annihilation operators involving also relativistic modes [Eq.~(\ref{Rindler_Bogoliubov_transformations})]. The non-relativistic limit in the inertial frame is non-equivalent to the non-relativistic limit in the accelerated frame. Moreover, the number of (anti)particle changes in the two frames.

We wonder if we can overcome such general differences in specific regimes. So far, we have considered an arbitrarily large acceleration. We may expect that in a limit in which the two frames are similar, the non-relativistic condition and the number of (anti)particles becomes approximately equivalent. In the following section we test the conditions for such equivalence to occur.

\section{Inertial and quasi-inertial frame}\label{Inertial_and_quasiinertial_frame}

In the present section, we consider the case in which the non-inertial observer has small acceleration with respect to the non-relativistic limit. Specifically, we require that
\begin{equation} \label{quasi_inertial_limit_a}
\frac{\hbar a}{m c} \sim \epsilon^{3/2}.
\end{equation}
The limits (\ref{non_relativistic_limit_curved}) and (\ref{quasi_inertial_limit_a}) can also be obtained by considering a diverging speed of light $c \rightarrow \infty$ with finite non-relativistic energy $E = \hbar \Omega - m c^2 \sim c^0$ and finite acceleration $\alpha = a c^2 \sim c^0$. We remark that Eq.~(\ref{quasi_inertial_limit_a}) is not a direct consequence of the non-relativistic limit, and it must be considered independently of Eq.~(\ref{non_relativistic_limit_curved}). Indeed, the limit $c \rightarrow \infty$ does not specify if $\alpha$ has to go to infinity with finite $a$, or $a$ has to go to zero with finite $\alpha$, or any other limiting scenarios.

The acceleration $a$ in Eq.~(\ref{quasi_inertial_limit_a}) is sufficiently high for non-inertial effects to be present in the non-relativistic limit. Indeed, when $a$ is such that Eq.~(\ref{quasi_inertial_limit_a}) holds, non-inertial corrections to the Hamiltonian are of the same order of non-relativistic energies \cite{falcone2022non}. Also, $a$ is low enough to preserve the non-relativistic condition and the number of particles, as we show in the present section.

In addition to Eq.~(\ref{quasi_inertial_limit_a}), we consider a further condition that defines the quasi-inertial limit. Specifically, we assume that quantum states are localized in a region of the right wedge such that
\begin{align} \label{quasi_inertial_limit_x}
& |ax - 1| \lesssim \epsilon, & a |X| \lesssim \epsilon.
\end{align}
Moreover, we assume that the non-inertial observer has only access to such region. For any $X$ such that Eq.~(\ref{quasi_inertial_limit_x}) holds, the metric $g_{\mu\nu}$ is approximated by $\eta_{\mu\nu}$ [Eq.~(\ref{Rindler_metric})]. This motivates our choice for the name \textit{quasi-inertial observer}.

The localization condition (\ref{quasi_inertial_limit_x}) defines the set of particles states that can be detected by the quasi-inertial observer. For instance, left-Rindler (anti)particles are excluded by such selection, since they are localized beyond the Rindler horizon. The same occurs for right-Rindler (anti)particles with frequency $\Omega \lesssim c a$, which are localized close to the horizon. Such localization is a consequence of the fact that the $F(\Omega,\vec{K}_\perp,T,\vec{X})$ modes are exponentially vanishing when $\Omega \lesssim c a$ and $a X \gtrsim -1$. One can see this by knowing that
\begin{equation}
K_{i \zeta} (\xi) \sim \frac{e^{-\xi}}{\sqrt{\xi}}
\end{equation}
when $\xi \rightarrow \infty$, and, hence, $F(\Omega,\vec{K}_\perp,T,\vec{X})$ is infinitesimal at least of order
\begin{equation} \label{F_Rindler_approximation_small_Omega}
F(\Omega,\vec{K}_\perp,T,\vec{X}) \sim \sqrt{\frac{\hbar}{a} \left| \sinh \left( \frac{\pi \Omega}{ca} \right) \right|} \epsilon^{3/4} \exp \left( - \epsilon^{-3/2} \right).
\end{equation}

We define a Fock space $\mathcal{F}_{\text{NQI}}$ that is generated by left-wedge (anti)particles with any frequency $\Omega$ and by right-wedge (anti)particles with frequency $\Omega \lesssim c a$. $\mathcal{F}_{\text{NQI}}$ represents the set of states that cannot be detected by the quasi-inertial observer. Therefore, we define the partial trace $\text{Tr}_{\text{NQI}}$ over $\mathcal{F}_{\text{NQI}}$. $\text{Tr}_{\text{NQI}}$ maps any pure state $| \Phi \rangle \in \mathcal{F}_\text{LR}$ into a statistical operator $\rho \in \mathcal{F}_\text{QI} = \text{Tr}_{\text{NQI}} \mathcal{F}_\text{LR}$ describing $| \Phi \rangle$ from the point of view of the non-inertial observer. In practice, the quasi-inertial observer is not able to distinguish between any element of $\mathcal{F}_{\text{NQI}}$ and the vacuum state of $\mathcal{F}_\text{QI}$.

In the following, we show that an inertial and a quasi-inertial observer agree about the first-quantization description of states that are localized in the region (\ref{quasi_inertial_limit_x}). Specifically, we prove that any localized non-relativistic Minkowski-Fock state $| \phi \rangle$ is also non-relativistic in the quasi-inertial frame, and that the number of (anti)particles and the wave functions are the same.

We start by clarifying what we mean by \textit{localized} Minkowski-Fock states with respect to Eq.~(\ref{quasi_inertial_limit_x}). Such localization condition is imposed on the wave functions of $|\phi \rangle$, which are defined in the following way
\begin{align} \label{free_wave_function}
\phi_{nm} ( \boldsymbol{x}) = & \left( \frac{2 m}{\hbar^2} \right)^{\frac{n+m}{2}} \int_{\mathbb{R}^{3(n+m)}} d^{3(n+m)} \boldsymbol{k} \tilde{\phi}_{nm} (\boldsymbol{k}) \nonumber \\
& \times \prod_{i=1}^n f(\vec{k}_i, 0, \vec{x}_i) \prod_{j=n+1}^{n+m} f(\vec{k}_j, 0, \vec{x}_j),
\end{align}
where
\begin{subequations}
\begin{align}
& \boldsymbol{x} = (\vec{x}_1, \dots, \vec{x}_n, \vec{x}_{n+1}, \dots, \vec{x}_{n+m}), \\ & \boldsymbol{k} = (\vec{k}_1, \dots, \vec{k}_n, \vec{k}_{n+1}, \dots, \vec{k}_{n+m})
\end{align}
\end{subequations}
are collections of $n+m$ vectors. $\tilde{\phi}_{nm} (\boldsymbol{k})$ is defined from the decomposition of $|\phi \rangle$ with respect to the Minkowski-Fock space $\mathcal{F}_\text{M}$
\begin{equation}\label{free_state_decomposition}
| \phi \rangle  = \hat{c}_\phi | 0_\text{M} \rangle,
\end{equation}
with
\begin{align}\label{c_phi}
\hat{c}_\phi  = & \sum_{n,m=0}^\infty \int_{\mathbb{R}^{3(n+m)}} d^{3(n+m)} \boldsymbol{k} \tilde{\phi}_{nm} (\boldsymbol{k})  \frac{1}{\sqrt{n!m!}} \prod_{i=1}^n \hat{a}^\dagger(\vec{k}_i)  \nonumber \\
& \times  \prod_{j=n+1}^{n+m} \hat{b}^\dagger(\vec{k}_j) .
\end{align}
$\tilde{\phi}_{nm} (\boldsymbol{k})$ is defined to be symmetric with respect to the momenta variables $\vec{k}_1, \dots, \vec{k}_n$ and with respect to $\vec{k}_{n+1}, \dots, \vec{k}_{n+m}$. Given the definition of wave functions for Minkowski states, one claims that $|\phi \rangle$ is localized in (\ref{quasi_inertial_limit_x}) if $\phi_{nm} ( \boldsymbol{x})$ is vanishing for any position variable $\vec{x}$ outside such region.

We consider a non-relativistic Minkowski-Fock state $| \phi \rangle \in \mathcal{F}_\text{M}$ that is localized in the region (\ref{quasi_inertial_limit_x}). We, hence, assume that $\tilde{\phi}_{nm} (\boldsymbol{k})$ and $\phi_{nm} ( \boldsymbol{x} )$ are non-vanishing when, respectively, all momenta are non-relativistic 
\begin{equation}\label{non_relativistic_momenta}
\frac{\hbar \omega(\vec{k})}{mc^2} - 1 \lesssim \epsilon
\end{equation}
and when all position variables are inside the region (\ref{quasi_inertial_limit_x}).

The explicit expression for $| \phi \rangle$ as an element of $\mathcal{F}_\text{LR}$ can be obtained from Eqs.~(\ref{Rindler_Bogoliubov_transformations_compact}), (\ref{Rindler_vacuum_to_Minkowski}), (\ref{Rindler_vacuum_to_Minkowski_unitary_operator}), (\ref{free_state_decomposition}), (\ref{c_phi}) and reads
\begin{equation}\label{free_state_decomposition_Rindler}
| \phi \rangle  = \hat{C}_\phi \hat{S} | 0_\text{L}, 0_\text{R} \rangle,
\end{equation}
with
\begin{align}\label{C_phi}
\hat{C}_\phi = & \sum_{n,m=0}^\infty \sum_{\boldsymbol{\nu}} \int_{\mathbb{R}^{3(n+m)}} d^{3(n+m)}  \boldsymbol{\Theta} \tilde{\Phi}_{nm} (\boldsymbol{\Theta}, \boldsymbol{\nu}) \frac{1}{\sqrt{n!m!}}  \nonumber \\
& \times\prod_{i=1}^n \hat{\mathcal{A}}^\dagger_{\nu_i}(\vec{\Theta}_i) \prod_{j=n+1}^{n+m} \hat{\mathcal{B}}^\dagger_{\nu_j}(\vec{\Theta}_j),
\end{align}
and
\begin{align}\label{non_reltaivistic_wavefunction_integration_0}
\tilde{\Phi}_{nm} (\boldsymbol{\Theta}, \boldsymbol{\nu}) = & \int_{\mathbb{R}^{3(n+m)}} d^{3(n+m)} \boldsymbol{k} \tilde{\phi}_{nm} (\boldsymbol{k}) \prod_{i=1}^n \alpha^*_{\nu_i}(\vec{k}_i,\vec{\Theta}_i) \nonumber \\
& \times \prod_{j=n+1}^{n+m} \alpha^*_{\nu_j}(\vec{k}_j,\vec{\Theta}_j),
\end{align}
where
\begin{subequations}
\begin{align}
& \boldsymbol{\Theta} = (\vec{\Theta}_1, \dots, \vec{\Theta}_n, \vec{\Theta}_{n+1}, \dots, \vec{\Theta}_{n+m}), \\ & \boldsymbol{\nu} = (\nu_1, \dots, \nu_n, \nu_{n+1}, \dots, \nu_{n+m})
\end{align}
\end{subequations}
are collections of $\vec{\Theta}$ and $\nu$ variables, and where the sum $\sum_{\boldsymbol{\nu}}$ in Eq.~(\ref{C_phi}) runs over all the possible $\nu$-variables $\nu \in \{ \text{L}, \text{R} \}$.

By using Eq.~(\ref{free_modes}) in Eq.~(\ref{alpha}) and by computing the derivative with respect to $\vec{x}_\perp$, one obtains
\begin{equation} \label{alpha_2}
\alpha_\nu(\vec{k},\vec{\Theta}) = \delta^2(\vec{k}_\perp-\vec{\Theta}_\perp) \chi_\nu (\vec{k},\Theta_1),
\end{equation}
with
\begin{align} \label{chi}
\chi_\nu(\vec{k},\Omega) = & \sqrt{ \frac{\pi}{\hbar c^2 \omega(\vec{k})} } \int_{\mathbb{R}} dx \theta(s_\nu x)   \left[  \frac{s_\nu \Omega}{a x} + \omega(\vec{k}) \right]  e^{-i k_1 x} \nonumber \\
& \times \tilde{F}(\Omega, \vec{k}_\perp, s_\nu X_\nu(x))
\end{align}
and $\vec{k}_\perp = (k_2, k_3)$ as transverse coordinates of momentum $\vec{k}$. As a consequence of the non-relativistic nature of $| \phi \rangle$ and thanks to the Dirac delta function appearing in Eq.~(\ref{alpha_2}), one deduces from Eq.~(\ref{non_reltaivistic_wavefunction_integration_0}) that $\tilde{\Phi}_{nm}$ is vanishing when at least one $\vec{\Theta}$-variable is such that $|\vec{\Theta}_\perp| \gg \epsilon^{1/2} mc / \hbar$. This leads to the following constrain for all $\vec{\Theta}$-variables
\begin{equation}\label{non_relativistic_Theta_perp}
 \frac{\hbar \Theta_\perp}{mc}  \lesssim  \epsilon^{1/2},
\end{equation}
which implies that each $\vec{\Theta}_\perp$ must be a non-relativistic momentum. 

Moreover, in the non-relativistic limit (\ref{non_relativistic_momenta}), $\alpha_\nu(\vec{k},\vec{\Theta})$ can be approximated by
\begin{equation} \label{alpha_alpha_tilde}
\alpha_\nu(\vec{k},\vec{\Theta}) \approx  \int_{\mathbb{R}^3} d^3 x f^*( \vec{k}, 0,  \vec{x} )  \tilde{\alpha}_\nu(\vec{x},\vec{\Theta}),
\end{equation}
with
\begin{align} \label{alpha_tilde}
\tilde{\alpha}_\nu(\vec{x},\vec{\Theta}) = & \frac{\theta(s_\nu x)}{\hbar c^2}  \left(  \frac{s_\nu \Theta_1}{a x} + \frac{m c^2}{\hbar} \right)  \tilde{F}(\vec{\Theta},  s_\nu X_\nu(x))  \nonumber \\
& \times e^{ i \vec{\Theta}_\perp \cdot \vec{x}_\perp}.
\end{align}
The relative error of Eq.~(\ref{alpha_alpha_tilde}) is of order $\epsilon$.

By using the relation between $\phi_{nm}$ and $\tilde{\phi}_{nm}$ [Eq.~(\ref{free_wave_function})] and between $\alpha_\nu(\vec{k},\vec{\Theta})$ and $\tilde{\alpha}_\nu(\vec{x},\vec{\Theta})$ [Eq.~(\ref{alpha_alpha_tilde})], one can approximate Eq. (\ref{non_reltaivistic_wavefunction_integration_0}) with
\begin{align}\label{non_reltaivistic_wavefunction_integration_0_tilde}
\tilde{\Phi}_{nm} (\boldsymbol{\Theta}, \boldsymbol{\nu}) \approx & \left( \frac{2 m}{\hbar^2} \right)^{-\frac{n+m}{2}} \int_{\mathbb{R}^{3(n+m)}} d^{3(n+m)} \boldsymbol{x} \phi_{nm} (\boldsymbol{x})  \nonumber \\
& \times \prod_{i=1}^n \tilde{\alpha}^*_{\nu_i}(\vec{x}_i,\vec{\Theta}_i) \prod_{j=n+1}^{n+m} \tilde{\alpha}^*_{\nu_j}(\vec{x}_j,\vec{\Theta}_j),
\end{align}
with relative error of order $\epsilon$. The locality condition can be used in Eq.~(\ref{non_reltaivistic_wavefunction_integration_0_tilde}) by recalling the fact that the wave function $\phi_{nm} ( \boldsymbol{x})$ is vanishing outside the region defined by $\vec{x}$-variables such that Eq.~(\ref{quasi_inertial_limit_x}) holds. The Heaviside theta function appearing in Eq.~(\ref{alpha_tilde}) implies that a necessary condition for the localization condition is that $\tilde{\Phi}_{nm} (\boldsymbol{\Theta}, \boldsymbol{\nu})$ is not vanishing only for all $\nu$ variables being equal to $\text{R}$. Therefore, hereafter we only consider the right-wedge wave function $\tilde{\Phi}_{nm} (\boldsymbol{\Theta})$ defined by
\begin{equation}
\tilde{\Phi}_{nm} (\boldsymbol{\Theta})= \tilde{\Phi}_{nm} (\boldsymbol{\Theta}, \textbf{R}),
\end{equation}
with
\begin{equation}
\boldsymbol{R} = (\underbrace{R, \dots, R}_n, \underbrace{R, \dots, R}_m).
\end{equation}

One may also introduce a cut-off $\delta x$ for any integration variable $x$ in Eq.~(\ref{non_reltaivistic_wavefunction_integration_0_tilde}) and assume that any integration can be approximately performed in $x \in [a^{-1}-\delta x, a^{-1}+\delta x]$ --- with $\delta x \sim \epsilon a^{-1}$ in the non-relativistic limit --- instead of the full real axis. By considering such approximation in Eq.~(\ref{non_reltaivistic_wavefunction_integration_0_tilde}) and using Eq.~(\ref{free_wave_function}), one obtains
\begin{align}\label{non_reltaivistic_wavefunction_integration_1}
\tilde{\Phi}_{nm} (\boldsymbol{\Theta} ) \approx & \int_{\mathbb{R}^{3(n+m)}} d^{3(n+m)} \boldsymbol{k} \tilde{\phi}_{nm} (\boldsymbol{k})  \prod_{i=1}^n \alpha^*(\vec{k}_i,\vec{\Theta}_i,\delta x) \nonumber \\
& \times \prod_{j=n+1}^{n+m} \alpha^*(\vec{k}_j,\vec{\Theta}_j,\delta x),
\end{align}
with
\begin{align} \label{alpha_bar}
\alpha(\vec{k},\vec{\Theta},\delta x) = & \frac{1}{\hbar c^2} \left(  \Theta_1 + \frac{m c^2}{\hbar} \right) \int_{a^{-1}-\delta x}^{a^{-1}+\delta x} d x \int_{\mathbb{R}^2} d^2 x_\perp   \nonumber \\
& \times f^*( \vec{k}, 0,\vec{x} )   \tilde{F}(\vec{\Theta},  X_\text{R}(x)) e^{ i \vec{\Theta}_\perp \cdot \vec{x}_\perp}.
\end{align}

By using Eq.~(\ref{free_modes}) and performing the integral with respect to $\vec{x}_\perp$, Eq.~(\ref{alpha_bar}) reads
\begin{equation} \label{alpha_bar_2}
\alpha(\vec{k},\vec{\Theta},\delta x) =  \delta^2(\vec{k}_\perp-\vec{\Theta}_\perp) \chi (\vec{k},\Theta_1, \delta x),
\end{equation}
with
\begin{align} \label{chi_bar_0}
\chi(\vec{k},\Omega,\delta x) = & \sqrt{ \frac{\pi}{\hbar c^2 \omega(\vec{k})} } \left(  \Omega + \frac{m c^2}{\hbar} \right)   \int_{a^{-1}-\delta x}^{a^{-1}+\delta x} d x  \nonumber \\
& \times e^{-i k_1 x}  \tilde{F}(\Omega,\vec{k}_\perp, X_\text{R}(x)),
\end{align}
which is the equivalent of Eq.~(\ref{chi}) with a cut-off $\delta x$ and $\omega(\vec{k}) \approx mc^2/\hbar$.

We are interested in the behavior of $\alpha(\vec{k},\vec{\Theta},\delta x)$ with varying $\Theta_1$ and we show that, when constraints (\ref{quasi_inertial_limit_a}), (\ref{quasi_inertial_limit_x}), (\ref{non_relativistic_momenta}) and (\ref{non_relativistic_Theta_perp}) hold, Eq.~(\ref{alpha_bar}) is not vanishing only for $\Theta_1$ such that
\begin{equation} \label{non_relativistic_Theta_1}
\left| \frac{\hbar \Theta_1}{m c^2} - 1 \right| \lesssim \epsilon.
\end{equation}
To this end, we perform the coordinate transformation
\begin{equation}\label{coordinate_transformation}
\bar{x} =  \frac{a x -1}{\bar{a}},
\end{equation}
with
\begin{equation}\label{acceleration_adimensional}
\bar{a} = 2^{-1/3} \left(\frac{\hbar a}{m c}\right)^{2/3}
\end{equation}
as acceleration-dependent adimensional variable. We furthermore consider the following adimensional variables
\begin{align}\label{adimensional_variables}
& \vec{\bar{k}} = \frac{\bar{a} \vec{k}}{a} , && \bar{\Omega} = \frac{\hbar \Theta_1}{m c^2} , && \delta \bar{x} = \frac{a \delta x}{\bar{a}}.
\end{align}
In this way, Eq.~(\ref{chi_bar_0}) reads
\begin{equation} \label{alpha_bar_chi_bar}
\chi(\vec{k},\vec{\Theta},\delta x) =  \frac{\bar{a}}{a}  \sqrt{\frac{ m \delta x}{\hbar}} \exp \left(- i \frac{k_1}{a} \right) \bar{\chi} \left(  \frac{\bar{a} \vec{k}}{a} , \frac{\hbar \Theta_1}{m c^2} , \frac{a \delta x}{\bar{a}} \right),
\end{equation}
with
\begin{equation} \label{chi_bar}
\bar{\chi}(\vec{\bar{k}}, \bar{\Omega}, \delta \bar{x}) =  \frac{\sqrt{\pi} (\bar{\Omega} + 1)}{\sqrt[4]{1 + 2 \bar{a} \bar{k}^2} \sqrt{\delta \bar{x}}}  \int_{- \delta \bar{x}}^{\delta \bar{x}} d \bar{x}  e^{ -i \bar{k}_1 \bar{x} } \bar{\tilde{F}} (\bar{\Omega}, \vec{\bar{k}}_\perp, \bar{x})
\end{equation}
and
\begin{equation} \label{F_bar}
\bar{\tilde{F}} (\bar{\Omega}, \vec{\bar{k}}_\perp, \bar{x}) = \sqrt{\frac{a}{\bar{a} \hbar}}   \tilde{F}\left( \frac{m c^2 \bar{\Omega}}{\hbar}, \frac{a \vec{\bar{k}}_\perp }{\bar{a}} ,  X_\text{R} \left( \frac{\bar{a} \bar{x} + 1}{a}\right) \right)
\end{equation}
as adimensional functions. The variable $ \vec{\bar{k}}_\perp$ appearing in Eq.~(\ref{chi_bar}) is made by the transverse components of $ \vec{\bar{k}}$, i.e.: $ \vec{\bar{k}}_\perp = (\bar{k}_2, \bar{k}_3)$.

\begin{figure}
\center
\includegraphics[]{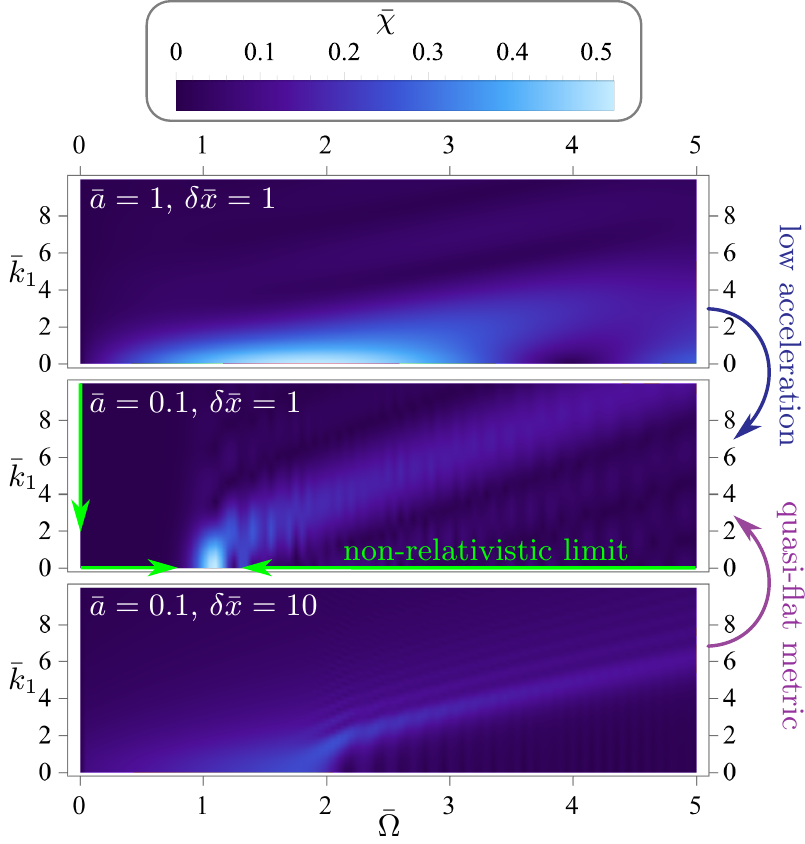}
\caption{Distribution of Rindler energies $\bar{\Omega}$ (horizontal axis) with respect to Minkowski momenta $\bar{k}_1$ (vertical axes). The quantity measured here is $\bar{\chi}(\vec{\bar{k}}, \bar{\Omega}, \delta \bar{x})$, which describes how energy-momentum wave functions transform from inertial to accelerated frames [Eqs.~(\ref{non_reltaivistic_wavefunction_integration_1}), (\ref{alpha_bar_2}), (\ref{alpha_bar_chi_bar})]. For simplicity, we ignore the transverse coordinates $y$ and $z$ by choosing $\bar{k}_2=0$ and $\bar{k}_3=0$. The regime of low acceleration ($\bar{a} \ll 1$) and quasi-flat metric ($\delta \bar{x} \sim 1$) [Eq.~(\ref{constraints_adimentional})] are indicated with, respectively, blue and purple arrows. In such regime, non-relativistic Minkowski momenta are paired with non-relativistic Rindler energies (green arrows). Indeed, when $\bar{k}_1 \lesssim 1$ [Eq.~(\ref{constraints_adimentional})], $\bar{\chi}(\vec{\bar{k}}, \bar{\Omega}, \delta \bar{x})$ is peaked for $\bar{\Omega} \approx 1$ [Eq.~(\ref{non_relativistic_Theta_1_adimensional})]. This means that in the quasi-inertial regime, the accelerated observer agrees with the inertial observer about the non-relativistic nature of particles.}\label{chi_bar_figure}
\end{figure}

Explicitly, Eq.~(\ref{chi_bar}) reads
\begin{align} \label{chi_bar_explicit}
 & \bar{\chi}(\vec{\bar{k}}, \bar{\Omega}, \delta \bar{x}) =   \frac{\bar{\Omega} + 1}{2 \pi^{3/2}\sqrt[4]{1 + 2 \bar{a} \bar{k}^2} \sqrt{\bar{a} \delta \bar{x}}} \int_{- \delta \bar{x}}^{\delta \bar{x}} d \bar{x} e^{ -i \bar{k}_1 \bar{x} }   \nonumber \\
 & \times \sqrt{ \left| \sinh \left( \frac{ \pi \bar{\Omega}}{ \sqrt{2 \bar{a}^3} } \right)\right| } K_{i \bar{\Omega} / \sqrt{2 \bar{a}^3} } \left( \sqrt{ \frac{1+ 2 \bar{a} \bar{k}_\perp^2 }{2 \bar{a}^3}} ( 1 + \bar{a} \bar{x} ) \right)
\end{align}
and gives the distribution of energies $\bar{\Omega}$ in the quasi-inertial frame for different $\vec{\bar{k}}$. In Fig.~\ref{chi_bar_figure} we plot such function for different values of $\bar{k}_1$ and $\bar{\Omega}$. We choose $\bar{a} \in \{0.1,1\}$ and $\delta \bar{x} \in \{ 1, 10 \}$ to show the quasi-inertial limit (i.e., $\bar{a} \ll 1$ and $\delta \bar{x} \lesssim 1$).

Conditions (\ref{quasi_inertial_limit_a}), (\ref{quasi_inertial_limit_x}), (\ref{non_relativistic_momenta}) in the new set of coordinates read
\begin{align}\label{constraints_adimentional}
& \bar{a} \sim \epsilon, && |\vec{\bar{k}}| \lesssim 1, && \delta \bar{x} \lesssim 1,
\end{align}
while Eq.~(\ref{non_relativistic_Theta_1}) reads
\begin{equation}\label{non_relativistic_Theta_1_adimensional}
\frac{|\bar{\Omega} - 1|}{\bar{a}} \lesssim 1.
\end{equation}
In \ref{appendix_b}, we show that when the coordinates $|\vec{\bar{k}}|$ and $\delta \bar{x}$ are constrained by Eq.~(\ref{constraints_adimentional}), $\bar{\chi}(\vec{\bar{k}}, \bar{\Omega}, \delta \bar{x})$ is not vanishing only for $\bar{\Omega}$ such that Eq.~(\ref{non_relativistic_Theta_1_adimensional}) holds. One can see this in Fig.~\ref{chi_bar_figure}, where the regime of low acceleration ($\bar{a} \ll 1$), quasi-flat metric ($\delta \bar{x} \lesssim 1$) and non-relativistic Minkowski momenta ($|\bar{k}_1| \lesssim 1$) is characterized by a distribution $\bar{\chi}(\vec{\bar{k}}, \bar{\Omega}, \delta \bar{x})$ peaked for non-relativistic Rindler energy ($\bar{\Omega} \sim 1$).

The result is that Eq.~(\ref{non_relativistic_Theta_1}), together with Eq.~(\ref{non_relativistic_Theta_perp}), selects the only $\vec{\Theta}$-variables for which $\tilde{\Phi}_{nm} (\boldsymbol{\Theta})$ is not vanishing. This means that $\hat{C}_\phi$ is approximately only made by creators and annihilators of non-relativistic right-Rindler particles. Therefore, the transformation $\hat{c}_\phi \mapsto \hat{C}_\phi$ conserves the non-relativistic nature of particles when one switches from the inertial to the accelerated frame.

Moreover, condition (\ref{non_relativistic_Theta_1}) implies that $\Theta_1>0$ and, hence, $\hat{\mathcal{A}}_\nu(\vec{\Theta})= \hat{A}_\nu(\vec{\Theta})$, $\hat{\mathcal{B}}_\nu(\vec{\Theta})= \hat{B}_\nu(\vec{\Theta})$. This leads to the following approximation for $\hat{C}_\phi$:
\begin{align}\label{C_phi_approximation}
\hat{C}_\phi \approx & \sum_{n,m=0}^\infty \int d^{3(n+m)}  \boldsymbol{\Theta} \tilde{\Phi}_{nm} (\boldsymbol{\Theta})  \frac{1}{\sqrt{n!m!}}  \prod_{i=1}^n \hat{A}^\dagger_\text{R}(\vec{\Theta}_i)  \nonumber \\
& \prod_{j=n+1}^{n+m} \hat{B}^\dagger_\text{R}(\vec{\Theta}_j).
\end{align}
Hereafter the integration intervals of $\boldsymbol{\Theta}$ are given by $\Theta_1 \in (0,\infty)$ and $\vec{\Theta}_\perp \in \mathbb{R}^2$ for each $\vec{\Theta}$-variable. Alternatively, one may use the intervals given by Eqs.~(\ref{non_relativistic_Theta_perp}) and (\ref{non_relativistic_Theta_1}), since, outside such region, $\tilde{\Phi}_{nm}$ vanishes.

By comparing Eq.~(\ref{C_phi_approximation}) with Eq.~(\ref{c_phi}) one can notice that $\hat{C}_\phi$ is identical to $\hat{c}_\phi$, up to the wave function $\tilde{\Phi}_{nm}$ replacing $\tilde{\phi}_{nm}$ and the right-Rindler creation operators $\hat{A}^\dagger_\text{R}$, $\hat{B}^\dagger_\text{R}$ replacing the Minkowski operators $\hat{a}^\dagger$, $\hat{b}^\dagger$. This implies that the number of particles and antiparticles created by $\hat{C}_\phi$ is the same of $\hat{c}_\phi$. The conclusion is that the transformation $\hat{c}_\phi \mapsto \hat{C}_\phi$ conserves the number of (anti)particles, in addition to the non-relativistic condition.

The approximation (\ref{C_phi_approximation}) can be used in Eq.~(\ref{free_state_decomposition_Rindler}) together with the following approximation for $\hat{S}$:
\begin{align}\label{Rindler_vacuum_to_Minkowski_approximation}
\hat{S} \approx &\exp\left(  2 \int_0^\Lambda d\Omega \int_{\mathbb{R}^2} d^2\vec{K}_\perp \exp \left( - \frac{\beta \Omega}{2} \right) \right. \nonumber \\
&  \times \left[ \hat{A}^\dagger_L(\Omega,\vec{K}_\perp) \hat{B}^\dagger_R(\Omega,-\vec{K}_\perp)  \right.\nonumber \\
& \left. \left. + \hat{B}^\dagger_L(\Omega,\vec{K}_\perp) \hat{A}^\dagger_R(\Omega,-\vec{K}_\perp) \right]^\text{A} \right) ,
\end{align}
where $\Lambda$ is a cut-off that excludes integration for $ \Omega \gg ca$. Equation (\ref{Rindler_vacuum_to_Minkowski_approximation}) can be derived from the fact that when $ \Omega \gg ca$, $\exp(-\beta \Omega / 2)$ is exponentially small.

One can notice that the integration interval in Eq.~(\ref{Rindler_vacuum_to_Minkowski_approximation}) is for $\Omega \lesssim ca \ll  m c^2 / \hbar $ [Eq.~(\ref{quasi_inertial_limit_a})], while the frequency variables $\Theta_1$ in Eq.~(\ref{C_phi_approximation}) are constrained by $\Theta_1 \approx mc^2/\hbar$ [Eq.~(\ref{non_relativistic_Theta_1})]. This means that $\hat{C}_\phi$ and $\hat{S}$ approximately commute:
\begin{equation} \label{C_phi_S_commutation_approximation}
[\hat{C}_\phi, \hat{S}] \approx 0.
\end{equation}
For the same reason, $\hat{C}_\phi$ is left unaffected by the partial trace $\text{Tr}_{\text{NQI}}$
\begin{equation}\label{partial_trace_C}
\text{Tr}_{\text{NQI}} ( \hat{C}_\phi \hat{O} ) \approx \hat{C}_\phi \text{Tr}_{\text{NQI}} ( \hat{O} ),
\end{equation}
while $\hat{S}$ satisfies the trace cyclic property
\begin{equation}\label{partial_trace_S}
\text{Tr}_{\text{NQI}} ( \hat{S} \hat{O} ) \approx \text{Tr}_{\text{NQI}} ( \hat{O} \hat{S} ).
\end{equation}
Indeed, the (anti)particles created by $\hat{C}_\phi$ do not belong to $\mathcal{F}_{\text{NQI}}$, since $\Theta_1 \approx mc^2/\hbar \gg ca$ [Eq.~(\ref{quasi_inertial_limit_a})]. On the other hand, (anti)particles created and annihilated by $\hat{S}$ have frequency $\Omega \lesssim ca$ and, hence, belong to $\mathcal{F}_{\text{NQI}}$.

Equations (\ref{partial_trace_C}) and (\ref{partial_trace_S}) can be used together with (\ref{free_state_decomposition_Rindler}) to prove that
\begin{equation} \label{partial_trace_phi}
 \text{Tr}_{\text{NQI}} ( | \phi \rangle \langle \phi | ) \approx \hat{C}_\phi  | 0_\text{QI} \rangle \langle  0_\text{QI} | \hat{C}^\dagger_\phi,
\end{equation}
where
\begin{equation}
| 0_\text{QI} \rangle \langle  0_\text{QI} | = \text{Tr}_{\text{NQI}} ( | 0_\text{L}, 0_\text{R} \rangle \langle 0_\text{L}, 0_\text{R} | )
\end{equation}
is the vacuum state of $\mathcal{F}_\text{QI}$. Equation (\ref{partial_trace_phi}) states that $| \phi \rangle$ is seen by the quasi-inertial observer through a pure state $| \Phi \rangle$ such that
\begin{equation} \label{Phi_C_phi}
| \Phi \rangle = \hat{C}_\phi  | 0_\text{QI} \rangle.
\end{equation}

In this way, we have proved that $| \phi \rangle$ is seen by the quasi-inertial observer as a non-relativistic state and with the same number of (anti)particles. Indeed, by comparing Eq.~(\ref{Phi_C_phi}) with Eq.~(\ref{free_state_decomposition}) and Eq.~(\ref{C_phi_approximation}) with Eq.~(\ref{c_phi}), one notices that the same number of non-relativistic (anti)particles are created over the respective vacuum. As said before, the map $\hat{c}_\phi \mapsto \hat{C}_\phi$ preserves the non-relativistic condition and the number of (anti)particles from the inertial to the quasi-inertial frame. The conclusion is that the inertial and the quasi-inertial observer agree about the first-quantization description of states.

Moreover, we have proved that $\tilde{\Phi}_{nm} (\boldsymbol{\Theta})$, defined by Eq.~(\ref{non_reltaivistic_wavefunction_integration_0}), plays the role of wave function of $| \Phi \rangle$ with respect to the quantum numbers $\boldsymbol{\Theta}$, analogously to $\tilde{\phi}_{nm}(\boldsymbol{k})$ in the inertial frame. The transformation $\tilde{\phi}_{nm} \mapsto \tilde{\Phi}_{nm}$ is given by Eq.~(\ref{non_reltaivistic_wavefunction_integration_0}).

The wave function of $| \Phi \rangle$ in the position representation, instead, can be defined by \cite{falcone2022non}
\begin{align} \label{free_wave_function_curved}
\Phi_{nm} ( \boldsymbol{X}) = & \left( \frac{2 m}{\hbar^2} \right)^{\frac{n+m}{2}} \int d^{3(n+m)} \boldsymbol{\Theta} \tilde{\Phi}_{nm} (\boldsymbol{\Theta})  \nonumber \\
& \times \prod_{i=1}^n F(\vec{\Theta}_i, 0,  \vec{X}_i)  \prod_{j=n+1}^{n+m} F(\vec{\Theta}_j, 0, \vec{X}_j).
\end{align}
The wave function transformation $\phi_{nm} \mapsto \Phi_{nm}$ can be derived by using Eq.~(\ref{non_reltaivistic_wavefunction_integration_0_tilde}) in Eq.~(\ref{free_wave_function_curved}):
\begin{align} \label{free_wave_function_curved_approx}
\Phi_{nm} ( \boldsymbol{X}) \approx & \int_{\mathbb{R}^{3(n+m)}} d^{3(n+m)} \boldsymbol{x} \phi_{nm} (\boldsymbol{x})   \prod_{i=1}^n \tilde{\tilde{\alpha}}^*_\text{R}(\vec{x}_i,\vec{X}_i) \nonumber \\
& \times \prod_{j=n+1}^{n+m} \tilde{\tilde{\alpha}}^*_\text{R}(\vec{x}_j,\vec{X}_j)
\end{align}
with
\begin{equation}
\tilde{\tilde{\alpha}}_\text{R}(\vec{x},\vec{X}) = \int_0^\infty d\Theta_1 \int_{\mathbb{R}^2} d\Theta_\perp \tilde{\alpha}_\text{R}(\vec{x},\vec{\Theta}) F^*(\vec{\Theta}, 0,  \vec{X}).
\end{equation}
As in Eq.~(\ref{non_reltaivistic_wavefunction_integration_0_tilde}), the relative error of Eq.~(\ref{free_wave_function_curved_approx}) is of order $\epsilon$.

In the non-relativistic (\ref{non_relativistic_momenta}), (\ref{non_relativistic_Theta_1}) and localized (\ref{quasi_inertial_limit_x}) limit, Eq.~(\ref{alpha_tilde}) can be approximated by
\begin{equation} \label{alpha_tilde_approx}
\tilde{\alpha}_\text{R}(\vec{x},\vec{\Theta}) \approx  \frac{2 \Theta_1}{\hbar c^2 a x}   \tilde{F}(\vec{\Theta}, X_\text{R}(x)) e^{i \vec{\Theta}_\perp \cdot \vec{x}_\perp},
\end{equation}
which leads to
\begin{align}\label{alpha_tilde_approx_2}
\tilde{\tilde{\alpha}}_\text{R}(\vec{x},\vec{X}) \approx &  \int_0^\infty d\Theta_1 \int_{\mathbb{R}^2} d\Theta_\perp \frac{2 \Theta_1}{\hbar c^2 a x}  \tilde{F}(\vec{\Theta}, X_\text{R}(x)) \nonumber \\
& \times \tilde{F}(\vec{\Theta}, X) e^{i \vec{\Theta}_\perp \cdot (\vec{x}_\perp - \vec{X}_\perp)}.
\end{align}
The relative error of Eq.~(\ref{alpha_tilde_approx}) is of order $\epsilon$.

It is possible to show that
\begin{align}\label{alpha_tilde_approx_3}
& \int_0^\infty d\Theta_1  \frac{2 \Theta_1}{\hbar c^2 a x} \tilde{F}(\vec{\Theta},X_\text{R}(x))  \tilde{F}(\vec{\Theta},X) \nonumber \\
 = & \frac{1}{4 \pi^2} \delta(x - x_\text{R}(X)),
\end{align}
where $x_\nu(X)$ is the inverse of Eq.~(\ref{Minkowski_to_Rindler_coordinates}), and, hence, the coordinate transformation (\ref{Rindler_coordinate_transformation_x}) with $t=T=0$:
\begin{equation}
a x_\nu = s_\nu \exp (s_\nu aX).
\end{equation}
A proof for Eq.~(\ref{alpha_tilde_approx_3}) is provided in \ref{appendix_2}.

Equations (\ref{alpha_tilde_approx_2}) and (\ref{alpha_tilde_approx_3}) lead to
\begin{equation}\label{alpha_tilde_approx_4}
\tilde{\tilde{\alpha}}_\text{R}(\vec{x},\vec{X}) \approx \delta(x - x_\text{R}(X)) \delta^2(\vec{x}_\perp - \vec{X}_\perp),
\end{equation}
which can be used in Eq.~(\ref{free_wave_function_curved_approx}) to obtain
\begin{equation} \label{free_wave_function_curved_approx_2}
\Phi_{nm} ( \boldsymbol{X}) \approx \phi_{nm} (\boldsymbol{x}_\text{R}(\boldsymbol{X})),
\end{equation}
where
\begin{align} \label{free_wave_function_curved_approx_2_x_nm}
& \boldsymbol{x}_\text{R}(\boldsymbol{X})  \nonumber \\
= & (\vec{x}_\text{R}(\vec{X}_1), \dots , \vec{x}_\text{R}(\vec{X}_n), \vec{x}_\text{R}(\vec{X}_{n+1}), \dots , \vec{x}_\text{R}(\vec{X}_{n+m})).
\end{align}
The function $\vec{x}_\nu(\vec{X})$ appearing in Eq.~(\ref{free_wave_function_curved_approx_2_x_nm}) is the coordinate transformation from the $\nu$-Rindler to the Minkowski spacetime when $t = T = 0$
\begin{equation}\label{Rindler_to_Minkowski_coordinate_t_T_0}
\vec{x}_\nu(\vec{X}) = (x_\nu(X), \vec{X}_\perp).
\end{equation}
Equation (\ref{free_wave_function_curved_approx_2}) states that the wave functions in the position representation approximately transform as scalars: $\Phi_{nm}$ is identical to $\phi_{nm}$ up to the coordinate transformation (\ref{Rindler_to_Minkowski_coordinate_t_T_0}) \footnote{Since the relative error of Eqs.~(\ref{free_wave_function_curved_approx}) and (\ref{alpha_tilde_approx}) is of order $\epsilon$, we can also infer that the relative error associated to the approximation (\ref{free_wave_function_curved_approx_2}) is of the same order. We remark that $\epsilon$ is also the order of the relative error associated to the inner product of states being approximated by the familiar $L^2(\mathbb{R}^3)$ inner product \cite{falcone2022non}. Therefore, the approximation (\ref{free_wave_function_curved_approx_2}) is compatible with the usual non-relativistic description of particles in terms of their wave functions in the position representation.}.

\section{Gaussian single-particle} \label{Gaussian_singleparticle}

We now provide an example of Minkowski single-particle state $| \phi \rangle$ to probe the results that we obtained. We assume that $\tilde{\phi}_{nm}$ is vanishing for any $n$ and $m$, except for $n=1$ and $m=0$. We also assume that the wave function $\tilde{\phi}_{10}(\vec{k})$ has a Gaussian form along the $x$ axis:
\begin{equation}\label{single_particle_Gaussian_Minkowski}
\tilde{\phi}_{10}(\vec{k}) = 2 \pi \tilde{\phi}(k_1) \delta(\vec{k}_\perp),
\end{equation}
with
\begin{equation}\label{single_particle_Gaussian_Minkowski_x}
\tilde{\phi}(k_1) = \frac{\sqrt{\sigma}}{\pi^{1/4} } \exp \left( - \frac{\sigma^2 k_1^2}{2} - i k_1 x_0 \right).
\end{equation}

In the position representation, the wave function $\phi_{10}$ [Eq.~(\ref{free_wave_function})] reads
\begin{equation}\label{single_particle_Gaussian_Minkowski_position}
\phi_{10}(\vec{x}) = \phi(x),
\end{equation}
with
\begin{equation}\label{single_particle_Gaussian_Minkowski_position_x}
\phi(x) = \frac{1}{\sqrt{2 \pi}} \int_{\mathbb{R}} dk_1 \sqrt{\frac{mc^2}{\hbar \omega(k_1 \vec{e}_1)}} \tilde{\phi}(k_1) e^{i k_1 x}
\end{equation}
and $\vec{e}_1 = (1,0,0)$. The non-relativistic limit leads to
\begin{equation}\label{single_particle_Gaussian_Minkowski_position_x_nonrelativistic}
\phi(x) \approx \frac{1}{\pi^{1/4} \sqrt{\sigma}} \exp \left( - \frac{ (x-x_0)^2}{2 \sigma^2} \right),
\end{equation}
which is a Gaussian wave function with variance $\sigma$.

Conversely, in the accelerated frame, the wave-functions $\tilde{\Phi}_{10}$ [Eqs.~(\ref{non_reltaivistic_wavefunction_integration_0})] and $\Phi_{10}$ [Eq.~(\ref{free_wave_function_curved})], respectively, read
\begin{align}
& \tilde{\Phi}_{10}(\Omega,\vec{K}_\perp) = 2 \pi \tilde{\Phi} (\Omega) \delta^2(\vec{K}_\perp),
& \Phi_{10} (\vec{X}) = \Phi (X),
\end{align}
with
\begin{subequations}\label{single_particle_Gaussian_Rindler_energy_position_x}
\begin{align}
& \tilde{\Phi} (\Omega) = \int_{\mathbb{R}} d k_1 \tilde{\phi}(k_1) \chi_\text{R}^*(k_1 \vec{e}_1,\Omega),\label{single_particle_Gaussian_Rindler_x} \\
& \Phi (X) = \frac{2 \pi \sqrt{2 m}}{\hbar} \int_0^\infty d\Omega \tilde{\Phi} (\Omega) \tilde{F}(\Omega \vec{e}_1,  X).\label{single_particle_Gaussian_Rindler_position_x}
\end{align}
\end{subequations}

In order for $| \phi \rangle$ to be non-relativistic in the inertial frame [Eq.~(\ref{non_relativistic_momenta})], we have to assume that
\begin{equation}\label{Gaussian_non_relativistic_0}
\frac{\hbar}{m c \sigma} \lesssim \epsilon^{1/2},
\end{equation}
which, together with Eq.~(\ref{quasi_inertial_limit_a}), reads
\begin{equation}\label{Gaussian_non_relativistic}
a \sigma  \gtrsim \epsilon.
\end{equation}
The localized condition (\ref{quasi_inertial_limit_x}), instead, requires
\begin{subequations}\label{Gaussian_local}
\begin{align}
& |ax_0 - 1| \lesssim \epsilon \label{Gaussian_local_x0} \\
 & a \sigma  \lesssim \epsilon \label{Gaussian_local_sigma}.
\end{align}
\end{subequations}

Hereafter we assume
\begin{align}\label{Gaussian_local_2}
 x_0  = \frac{1}{a},
\end{align}
in order to meet condition (\ref{Gaussian_local_x0}). On the other hand, we consider different values of $\sigma$, which are constrained by Eqs.~(\ref{Gaussian_non_relativistic}) and (\ref{Gaussian_local_sigma}):
\begin{equation}\label{Gaussian_non_relativistic_local_sigma}
a \sigma  \sim \epsilon.
\end{equation}

We consider the adimensional variables defined by Eqs.~(\ref{coordinate_transformation}), (\ref{acceleration_adimensional}), (\ref{adimensional_variables}), together with
\begin{align}
& \bar{\sigma} = \frac{a \sigma}{\bar{a}}, & \bar{X} = \frac{a X}{\bar{a}}
\end{align}
and the following adimensional wave functions
\begin{subequations}
\begin{align}
& \bar{\tilde{\phi}}(\bar{k}_1) =  \sqrt{\frac{a}{\bar{a}}} \exp \left(  i \frac{\bar{k}_1}{\bar{a}} \right) \tilde{\phi} \left( \frac{a \bar{k}_1}{\bar{a}} \right), \\
& \bar{\phi}(\bar{x}) =  \sqrt{\frac{\bar{a}}{a}} \phi \left( \frac{\bar{a} \bar{x} + 1}{a} \right),\\
& \bar{\tilde{\Phi}}(\bar{\Omega}) =  \sqrt{\frac{m c^2}{\hbar}} \tilde{\Phi} \left( \frac{mc^2 \bar{\Omega}}{\hbar} \right), \\
 & \bar{\Phi}(\bar{X}) =  \sqrt{\frac{\bar{a}}{a}}  \Phi \left( \frac{\bar{a} \bar{X}}{a} \right).
\end{align}
\end{subequations}

In this way Eqs.~(\ref{single_particle_Gaussian_Minkowski_x}), (\ref{single_particle_Gaussian_Minkowski_position_x}), (\ref{single_particle_Gaussian_Rindler_energy_position_x}) read
\begin{subequations}\label{single_particle_x_adimensional}
\begin{align}
\bar{\tilde{\phi}}(\bar{k}_1) = & \frac{\sqrt{\bar{\sigma}}}{\pi^{1/4} } \exp \left( - \frac{\bar{\sigma}^2 \bar{k}_1^2}{2}  \right),\label{single_particle_Gaussian_Minkowski_x_adimensional}\\
\bar{\phi}(\bar{x}) = & \frac{1}{\sqrt{2 \pi}} \int_{\mathbb{R}} d\bar{k}_1 \frac{e^{i \bar{k}_1 \bar{x}} \bar{\tilde{\phi}}(\bar{k}_1)}{\sqrt[4]{ 1 + 2 \bar{a} \bar{k}_1^2 }} ,\label{single_particle_Gaussian_Minkowski_position_x_adimensional}\\
\bar{\tilde{\Phi}} (\bar{\Omega}) = & \frac{1}{\sqrt{\bar{a}}} \int_{\mathbb{R}} d \bar{k}_1   \bar{\tilde{\phi}}(\bar{k}_1) \bar{\chi}_\text{R}^*(\bar{k}_1 \vec{e}_1,\bar{\Omega}), \label{single_particle_Gaussian_Rindler_x_adimensional}\\
\bar{\Phi} (\bar{X}) = &  \frac{2 \pi}{\sqrt{\bar{a}}} \int_0^\infty d\bar{\Omega} \bar{\tilde{\Phi}} (\bar{\Omega})  \bar{\tilde{F}}( \bar{\Omega}  \vec{e}_1, \bar{x}_\text{R}(\bar{X}) ), \label{single_particle_Gaussian_Rindler_position_x_adimensional}
\end{align}
\end{subequations}
where
\begin{equation}\label{coordinate_transformation_adimensional}
\bar{x}_\text{R}(\bar{X}) = \frac{1}{\bar{a}} \left[ a x_\text{R} \left( \frac{\bar{a} \bar{X}}{a} \right) - 1\right]
\end{equation}
is the coordinate transformation between the adimensional variables $\bar{x}$ and $\bar{X}$, and where $\bar{\chi}_\nu$ is defined as the adimensional equivalent of $\chi_\nu(\vec{k},\Omega)$ by the following identity
\begin{equation} \label{alpha_bar_chi_bar_nu}
\chi_\nu(\vec{k},\Omega) = \sqrt{\frac{\hbar}{m c^2 a }}  \exp \left(- i \frac{k_1}{a} \right) \bar{\chi}_\nu \left(  \frac{\bar{a} \vec{k}}{a}  , \frac{\hbar \Omega}{m c^2}  \right).
\end{equation}
Moreover, condition (\ref{Gaussian_non_relativistic_local_sigma}) now reads
\begin{equation}\label{Gaussian_non_relativistic_local_sigma_adimensional}
\bar{\sigma}  \sim 1.
\end{equation}

The explicit form of $\bar{\chi}_\text{R}(\bar{k}_1 \vec{e}_1,\bar{\Omega})$ appearing in Eq.~(\ref{single_particle_Gaussian_Rindler_x_adimensional}) can be obtained by performing the integral in Eq.~(\ref{chi}), which leads to \cite{RevModPhys.80.787}
\begin{align} \label{chi_nu_explicit}
& \chi_\text{R}(\vec{k},\Omega) = \left[ 4 \pi a \omega(k) \left| \sinh \left( \frac{\beta \Omega}{2} \right) \right| \right]^{-1/2}  \nonumber \\
& \times \exp \left( \frac{\beta \Omega}{4}  - i \frac{\Omega}{2 c a} \ln \left( \frac{\omega(\vec{k})+ck_1}{\omega(\vec{k})-ck_1} \right) \right).
\end{align}
The adimensional equivalent of Eq.~(\ref{chi_nu_explicit}) reads
\begin{align} \label{chi_nu_bar_explicit}
& \bar{\chi}_\text{R}(\vec{\bar{k}},\bar{\Omega}) = \left[ 4 \pi  \sqrt{1+2 \bar{a} \bar{k}^2} \left| \sinh \left( \frac{\pi \bar{\Omega}}{\sqrt{2 \bar{a}^3}} \right) \right| \right]^{-1/2} \nonumber \\
& \times \exp \left( \frac{ \pi \bar{\Omega}}{(2 \bar{a})^{3/2}}  + i \frac{\bar{k}_1}{\bar{a}} \right. \nonumber \\
& \left. - i \frac{ \bar{\Omega}}{(2 \bar{a})^{3/2}} \ln \left( \frac{\sqrt{1+2 \bar{a} \bar{k}^2}+ \sqrt{2 \bar{a}} \bar{k}_1 }{\sqrt{1+2 \bar{a} \bar{k}^2}-\sqrt{2 \bar{a}} \bar{k}_1} \right) \right),
\end{align}
and can be used in Eq.~(\ref{single_particle_Gaussian_Rindler_x_adimensional}) to derive the explicit form of $\bar{\tilde{\Phi}} (\bar{\Omega})$.

\begin{figure}
\includegraphics[scale=0.95]{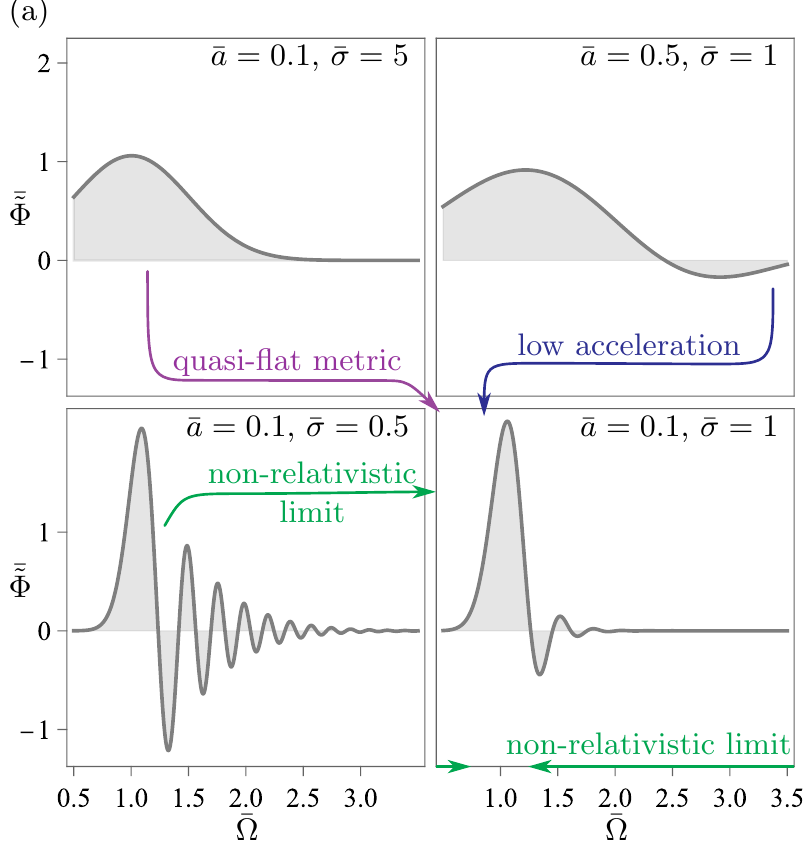}
\includegraphics[scale=0.95]{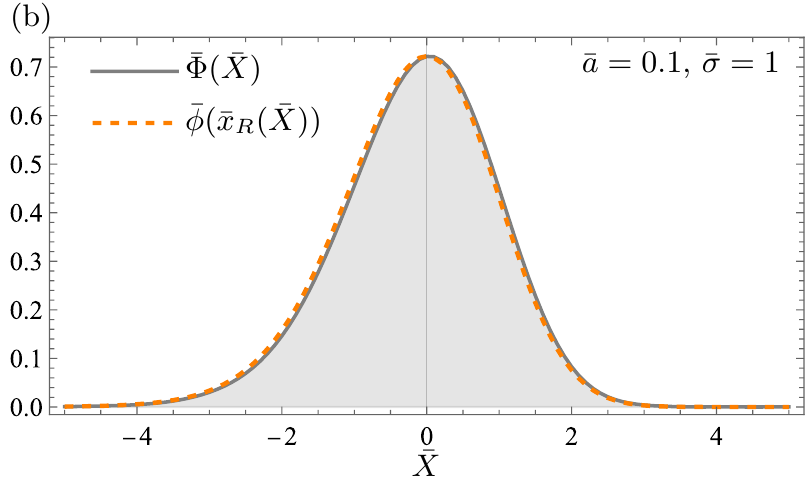}
\caption{Inertial Gaussian single-particle wave functions in accelerated frames. In panel (a), we plot the distribution of Rindler frequencies $\bar{\Omega}$ with respect to different acceleration $\bar{a}$ and different variance $\bar{\sigma}$. If $\bar{a} = 0.1$, $\bar{\sigma} = 1$, the wave function $\bar{\tilde{\Phi}}(\bar{\Omega})$ is peaked in $\bar{\Omega} \approx 1$ and, hence, the state is populated by non-relativistic energies in the accelerated frame [Eq.~(\ref{non_relativistic_Theta_1_adimensional})]. Conversely, relativistic energies appear for other configurations. The reasons are the following: when $\bar{\sigma} = 5$, the particle is not well-localized in the region where the metric is almost flat [Eq.~(\ref{Gaussian_local_sigma})]; when $\bar{\sigma} = 0.5$ the state is populated by relativistic Minkowski momenta [Eq.~(\ref{Gaussian_non_relativistic})]; when $\bar{a}=0.5$ the acceleration is not sufficiently low for the quasi-inertial approximation [Eq.~(\ref{quasi_inertial_limit_a})]. In panel (b), we show the wave function in the position representation $\bar{\Phi}(\bar{X})$ (gray solid line) for the state seen by the accelerated observer. We chose $\bar{a} = 0.1$ and $\bar{\sigma} = 1$ for the non-relativistic and quasi-inertial approximation. In such regime, $\bar{\Phi}(\bar{X})$ can be approximated by the Minkowski wave function $\bar{\phi}(\bar{x}_\text{R}(\bar{X}))$ (orange dashed line) under the coordinate transformation $\bar{x}_\text{R}(\bar{X})$.}\label{Gaussian_figure}
\end{figure}

By using Eqs.~(\ref{F_tilde_Rindler}), (\ref{F_bar}), (\ref{chi_nu_bar_explicit}) in Eq.~(\ref{single_particle_x_adimensional}), one is able to compute the wave functions $\bar{\tilde{\Phi}} (\bar{\Omega})$ and $\bar{\Phi} (\bar{X})$ in the accelerated frame. The results are drawn in Fig.~\ref{Gaussian_figure}.

In Fig.~\ref{Gaussian_figure}a, we show that under condition (\ref{Gaussian_non_relativistic_local_sigma_adimensional}) and $\bar{a} \ll 1$, $\bar{\tilde{\Phi}} (\bar{\Omega})$ is not vanishing only for non-relativistic frequencies ($\bar{\Omega} \approx 1$). This is in agreement with the results of Sec.~\ref{Inertial_and_quasiinertial_frame}: in the quasi-inertial regime ($\bar{\sigma} \lesssim 1$, $\bar{a} \ll 1$), the accelerated observer detects non-relativistic particles ($\bar{\Omega} \sim 1$) when the state is non-relativistic also in the inertial frame ($\bar{\sigma} \gtrsim 1$). Conversely, when conditions (\ref{Gaussian_non_relativistic_local_sigma_adimensional}) and $\bar{a} \ll 1$ are not met, relativistic energies are present in the accelerated frame.

In Fig.~\ref{Gaussian_figure}b, we plot the wave function $\bar{\Phi} (\bar{X})$. We choose a configuration in which condition (\ref{Gaussian_non_relativistic_local_sigma_adimensional}) and $\bar{a} \ll 1$ are met. One can see that in such case, $\bar{\Phi} (\bar{X})$ is approximated by $\bar{\phi}(\bar{x})$, up to the coordinate transformation (\ref{coordinate_transformation_adimensional}). Such result confirms the prediction of Eq.~(\ref{free_wave_function_curved_approx_2}) for the case of a single Gaussian particle.

\section{Conclusions} \label{Conclusions}

We have shown the frame-dependence of the non-relativistic limit. Specifically, we have shown that by switching from an inertial to a non-inertial frame, the relativistic nature of quantum states may change: non-relativistic particles of one frame can be relativistic for the other observer. Also the number of particles may change. 

This can be problematic in the context of non-inertial detectors --- e.g., Unruh-DeWitt detectors \cite{PhysRevD.14.870, PhysRevD.29.1047, hawking1980general}. For instance, an atomic detector --- that is prepared in the laboratory frame as a non-relativistic $n$-particles state and then accelerated --- cannot be described as a fixed number of non-relativistic particles in its proper non-inertial frame. The familiar first-quantization description of atomic systems breaks down when one switches from the inertial to the accelerated frame.

We have proposed a solution to such problem by considering a quasi-inertial frame. The observer is defined to have low acceleration in the non-relativistic limit --- but high enough to see non-inertial effects --- and can only have access to a region in which the metric is quasi-flat. We have shown that non-relativistic states in the inertial frame are also non-relativistic in the quasi-inertial frame, as opposed to the case of arbitrarily large accelerations. Moreover, the number of particles is preserved when switching from one frame to the other. This provides a solution to the problems mentioned above.

Also, we have shown how particles wave functions transform from the inertial to the quasi-inertial frame. Specifically, we have proved that such functions approximately transform as scalar fields under the coordinate transformation.

We believe that these results may be useful in future works about non-relativistic particles seen by inertial and non-inertial observers, such as accelerated Unruh-DeWitt detectors.

\appendix
\section{}\label{appendix_a}

We prove Eqs.~(\ref{Rindler_Bogoliubov_transformations}) and (\ref{alpha}). We use the procedure shown in Sec.~\ref{Inertial_and_non_inertial_frame} that led to Eq.~(\ref{Bogoliubov_transformations}) through Eqs.~(\ref{scalar_transformation}), (\ref{Klein_Gordon_curved_Phi}) and (\ref{Bogoliubov_transformations_1}).

An explicit decomposition of the field in Minkowski spacetime is given by Eq.~(\ref{free_field}). Therefore, the equivalent of Eq.~(\ref{Bogoliubov_transformations_1}) reads
\begin{align}\label{Bogoliubov_transformations_1_Rindler}
& \hat{a}(\vec{k}) = ( f(\vec{k}), \hat{\phi})_\text{KG}, 
& \hat{b}^\dagger(\vec{k}) = -(f^*(\vec{k}),  \hat{\phi})_\text{KG},
\end{align}
which explicitly reads
\begin{subequations}\label{Bogoliubov_transformations_2_Rindler}
\begin{align}
\hat{a}(\vec{k}) = & \frac{i}{\hbar c^2} \int_{\mathbb{R}^3} d^3x \left[  f^*(\vec{k},t,\vec{x}) \partial_0 \hat{\phi}(t,\vec{x}) \right. \nonumber \\
& \left. - \hat{\phi}(t,\vec{x}) \partial_0  f^*(\vec{k},t,\vec{x})\right], \\
\hat{b}^\dagger(\vec{k}) = & - \frac{i}{\hbar c^2} \int_{\mathbb{R}^3} d^3x \left[  f(\vec{k},t,\vec{x}) \partial_0 \hat{\phi}(t,\vec{x})  \right. \nonumber \\
& \left. - \hat{\phi}(t,\vec{x}) \partial_0  f(\vec{k},t,\vec{x})\right].
\end{align}
\end{subequations}

The transformation between fields $\hat{\Phi}_\nu \mapsto \hat{\phi}$ when $t=T=0$ is given by the inverse of Eq.~(\ref{scalar_transformation_Rindler}):
\begin{equation}\label{scalar_transformation_Rindler_inverse}
\hat{\phi}(0,\vec{x}) = \begin{cases}
\hat{\Phi}_\text{L}(0,\vec{X}_\text{L}(\vec{x})) & \text{if } x<0\\
\hat{\Phi}_\text{R}(0,\vec{X}_\text{R}(\vec{x})) & \text{if } x>0
\end{cases},
\end{equation}
where $\vec{X}_\nu(\vec{x})$ is the coordinate transformation from the Minkowski to the $\nu$-Rindler spacetime when $t = T = 0$:
\begin{equation}
\vec{X}_\nu(\vec{x}) = (X_\nu(x), \vec{x}_\perp).
\end{equation}
Analogously, $\partial_0 \hat{\phi}(t,\vec{x})$ can be obtained from $\hat{\Phi}_\nu(T,\vec{X})$ by considering the following chain rule:
\begin{equation}
\frac{\partial}{\partial T} = \frac{\partial t_\nu}{\partial T}  \frac{\partial}{\partial t}  + \frac{\partial x_\nu}{\partial T}  \frac{\partial}{\partial x} + \frac{\partial y_\nu}{\partial T}  \frac{\partial}{\partial y} + \frac{\partial z_\nu}{\partial T}  \frac{\partial}{\partial z}, 
\end{equation}
which in the Rindler spacetime reads
\begin{equation}\label{chain_rule_Rindler}
\frac{\partial}{\partial T} = s_\nu ax \frac{\partial}{\partial t}  + s_\nu ac^2t  \frac{\partial}{\partial x}.
\end{equation}
By using Eq.~(\ref{chain_rule_Rindler}) in Eq.~(\ref{scalar_transformation_Rindler}) and choosing $t=T=0$, one obtains
\begin{equation}\label{time_derivative_scalar_transformation_Rindler}
\partial_0 \hat{\phi}(0,\vec{x}) = \begin{cases}
-(ax)^{-1} \partial_0 \hat{\Phi}_\text{L}(0,\vec{X}_\text{L}(\vec{x})) & \text{if } x<0\\
(ax)^{-1} \partial_0 \hat{\Phi}_\text{R}(0,\vec{X}_\text{R}(\vec{x})) & \text{if } x>0
\end{cases}.
\end{equation}

In a more compact way, Eqs.~(\ref{scalar_transformation_Rindler_inverse}) and (\ref{time_derivative_scalar_transformation_Rindler}) read, respectively,
\begin{equation}\label{scalar_transformation_Rindler_inverse_2}
\hat{\phi}(0,\vec{x}) = \sum_{\nu=\{\text{L},\text{R}\}} \theta(s_\nu x) \hat{\Phi}_\nu(0,\vec{X}_\nu(\vec{x}))
\end{equation}
and
\begin{equation}\label{time_derivative_scalar_transformation_Rindler_2}
\partial_0 \hat{\phi}(0,\vec{x}) = \sum_{\nu=\{\text{L},\text{R}\}} \theta(s_\nu x) \frac{s_\nu}{a x} \partial_0 \hat{\Phi}_\nu(0,\vec{X}_\nu(\vec{x})).
\end{equation}

By choosing $t=0$ and using Eqs.~(\ref{scalar_transformation_Rindler_inverse_2}) and (\ref{time_derivative_scalar_transformation_Rindler_2}) in Eq.~(\ref{Bogoliubov_transformations_2_Rindler}) one obtains
\begin{subequations}\label{Bogoliubov_transformations_3_Rindler}
\begin{align}
\hat{a}(\vec{k}) = & \frac{i}{\hbar c^2} \sum_{\nu=\{\text{L},\text{R}\}} \int_{\mathbb{R}^3} d^3x \theta(s_\nu x)  \nonumber \\
& \times\left[ \frac{s_\nu}{a x} f^*(\vec{k},0,\vec{x}) \partial_0 \hat{\Phi}_\nu(0,\vec{X}_\nu(\vec{x}))  \right. \nonumber \\
& \left. - \hat{\Phi}_\nu(0,\vec{X}_\nu(\vec{x})) \partial_0  f^*(\vec{k},0,\vec{x})  \right], \\
\hat{b}^\dagger(\vec{k}) = & -\frac{i}{\hbar c^2} \sum_{\nu=\{\text{L},\text{R}\}} \int_{\mathbb{R}^3} d^3x \theta(s_\nu x)  \nonumber \\
& \times \left[ \frac{s_\nu}{a x} f(\vec{k},0,\vec{x}) \partial_0 \hat{\Phi}_\nu(0,\vec{X}_\nu(\vec{x}))  \right. \nonumber \\
& \left. - \hat{\Phi}_\nu(0,\vec{X}_\nu(\vec{x})) \partial_0  f(\vec{k},0,\vec{x})  \right].
\end{align}
\end{subequations}
Equation (\ref{Rindler_scalar_decomposition}) can be used in Eq.~(\ref{Bogoliubov_transformations_3_Rindler}) to obtain
\begin{subequations}\label{Rindler_Bogoliubov_transformations_1}
\begin{align}
\hat{a}(\vec{\Theta}) = & \sum_{\nu=\{\text{L},\text{R}\}} \int_0^\infty d\Theta_1 \int_{\mathbb{R}^2} d^2\Theta_\perp \left[ \alpha_{\nu+}(\vec{k},\vec{\Theta}) \hat{A}_\nu(\vec{\Theta}) \right. \nonumber \\
& \left. + \alpha_{\nu-}(\vec{k},\vec{\Theta}) \hat{B}_\nu^\dagger(\vec{\Theta}) \right],  \\
\hat{b}^\dagger(\vec{\Theta}) = & \sum_{\nu=\{\text{L},\text{R}\}} \int_0^\infty d\Theta_1 \int_{\mathbb{R}^2} d^2\Theta_\perp \left[ \alpha^*_{\nu+}(\vec{k},\vec{\Theta}) \hat{B}_\nu^\dagger(\vec{\Theta}) \right. \nonumber \\
& \left.+ \alpha^*_{\nu-}(\vec{k},\vec{\Theta}) \hat{A}_\nu(\vec{\Theta}) \right],
\end{align}
\end{subequations}
with 
\begin{subequations}\label{alpha_nu_pm}
\begin{align} 
\alpha_{\nu+}(\vec{k},\vec{\Theta}) = &   \frac{i}{\hbar c^2} \int_{\mathbb{R}^3} d^3x \theta(s_\nu x)   \nonumber \\
& \times \left[ \frac{s_\nu}{a x} f^*(\vec{k},0,\vec{x}) \partial_0 F(\vec{\Theta}, 0,s_\nu \vec{X}_\nu(\vec{x})) \right. \nonumber \\
& \left. - F(\vec{\Theta}, 0,s_\nu \vec{X}_\nu(\vec{x})) \partial_0  f^*(\vec{k},0,\vec{x})  \right],\\
\alpha_{\nu-}(\vec{k},\vec{\Theta}) =  & \frac{i}{\hbar c^2} \int_{\mathbb{R}^3} d^3x \theta(s_\nu x) \nonumber \\
& \times \left[ \frac{s_\nu}{a x} f^*(\vec{k},0,\vec{x}) \partial_0 F^*(\vec{\Theta}, 0,s_\nu \vec{X}_\nu(\vec{x}))   \right. \nonumber \\
& \left. - F^*(\vec{\Theta}, 0,s_\nu \vec{X}_\nu(\vec{x})) \partial_0  f^*(\vec{k},0,\vec{x})  \right].
\end{align}
\end{subequations}

By using Eqs.~(\ref{free_modes}), (\ref{F_Rindler}) and the fact that $\tilde{F}$ is real, Eq.~(\ref{alpha_nu_pm}) reads
\begin{align}\label{alpha_nu_pm_2}
\alpha_{\nu \pm}(\vec{k},\vec{\Theta}) = & \frac{1}{\hbar c^2} \int_{\mathbb{R}^3} d^3 x \theta(s_\nu x) \left[ \pm  \frac{s_\nu \Theta_1}{a x} + \omega(\vec{k}) \right]  \nonumber \\
& \times f^*( \vec{k}, 0, \vec{x} )  \tilde{F}(\vec{\Theta},  s_\nu X_\nu(x)) e^{ \pm i \vec{\Theta}_\perp \cdot \vec{x}_\perp}.
\end{align}
By knowing that $\tilde{F}(\vec{\Theta}, X)$ is invariant under $\vec{\Theta} \mapsto - \vec{\Theta}$ [Eq.~(\ref{F_tilde_Rindler})], Eq.~(\ref{alpha_nu_pm_2}) reads
\begin{equation}\label{alpha_nu_pm_4}
\alpha_{\nu \pm}(\vec{k},\vec{\Theta}) = \alpha_\nu (\vec{k},\pm \vec{\Theta}),
\end{equation}
with $\alpha_\nu$ defined by Eq.~(\ref{alpha}).

Equations (\ref{Rindler_Bogoliubov_transformations_1}) and (\ref{alpha_nu_pm_4}) finally prove Eq.~(\ref{Rindler_Bogoliubov_transformations}).

\section{} \label{appendix_b}

We assume that Eq.~(\ref{constraints_adimentional}) holds. We prove that $\bar{\chi}(\vec{\bar{k}}, \bar{\Omega}, \delta \bar{x})$ is vanishing when $|\bar{\Omega} - 1| \gg \bar{a}$. We start by considering the case $|\bar{\Omega}| \lesssim  \bar{a}^{3/2}$, which is a sufficient condition for $|\bar{\Omega} - 1| \gg \bar{a}$. The limit $|\bar{\Omega}| \lesssim  \bar{a}^{3/2}$ is equivalent to $|\Theta_1| \lesssim ca$, and, hence, leads to exponentially vanishing Rindler modes $\bar{\tilde{F}} (\bar{\Omega}, \vec{\bar{k}}_\perp, \bar{x})$ appearing in Eq.~(\ref{chi_bar}), as we have already shown in Eq.~(\ref{F_Rindler_approximation_small_Omega}).

Conversely, if $|\bar{\Omega}| \gg  \bar{a}^{3/2}$, $\bar{\tilde{F}} (\bar{\Omega}, \vec{\bar{k}}_\perp, \bar{x})$ can be approximated by \cite{Dunster1990BesselFO, olver2014asymptotics}
\begin{align}\label{F_tilde_Rindler_Hankel_2}
\bar{\tilde{F}} (\bar{\Omega},\vec{\bar{k}}_\perp, \bar{x})  \approx & \frac{1}{2 \pi} \sqrt[6]{\frac{2}{\bar{\Omega}^2}} \sqrt[4]{\frac{\zeta(z(\bar{\Omega},\vec{\bar{k}}_\perp,\bar{x}))}{1-z^2(\bar{\Omega},\vec{\bar{k}}_\perp,\bar{x})} }  \nonumber \\
& \times \text{Ai} \left( - \frac{1}{ \bar{a}} \sqrt[3]{\frac{\bar{\Omega}^2}{2}} \zeta(z(\bar{\Omega},\vec{\bar{k}}_\perp,\bar{x})) \right),
\end{align}
with
\begin{subequations}\label{z_zeta_z}
\begin{equation}
z(\bar{\Omega},\vec{\bar{k}}_\perp,\bar{x}) = \frac{1}{|\bar{\Omega}|} \sqrt{ 1+ 2 \bar{a} \bar{k}_\perp^2 }  (1 + \bar{a} \bar{x}) ,
\end{equation}
\begin{equation}\label{zeta_z}
\begin{cases}
\frac{2}{3} \zeta^{3/2}(z) = \ln \left( \frac{1+\sqrt{1-z^2}}{z} \right) - \sqrt{1-z^2}, & \text{if } 0 \leq z \leq 1,  \\
\frac{2}{3} [-\zeta(z)]^{3/2} = \sqrt{z^2-1} - \arccos \left( \frac{1}{z} \right), & \text{if }  z \geq 1 
\end{cases}
\end{equation}
\end{subequations}
and where $\text{Ai}(\xi)$ is the Airy function.

When conditions (\ref{constraints_adimentional}) hold and when $||\tilde{\Omega}|-1| \sim \bar{a}$, the variables $z(\bar{\Omega},\vec{\bar{k}}_\perp,\bar{x})$ and $\zeta(z)$ can be approximated by the following expansion \cite{olver2014asymptotics}
\begin{subequations}\label{z_zeta_z_approximation_Puiseux}
\begin{align}
& z(\bar{\Omega},\vec{\bar{k}}_\perp,\bar{x}) \approx  1  +  \bar{a} (\bar{k}_\perp^2 + \bar{x}) - (|\bar{\Omega}|  -1) ,\\
&   \zeta(z(\bar{\Omega},\vec{\bar{k}}_\perp,\bar{x})) \approx -\sqrt[3]{2} [ \bar{a} (\bar{k}_\perp^2 + \bar{x}) - (|\bar{\Omega}|  -1) ]. \label{zeta_z_approximation_Puiseux}
\end{align}
\end{subequations}
If $||\tilde{\Omega}|-1| \gg \bar{a}$, instead, $z(\bar{\Omega},\vec{\bar{k}}_\perp,\bar{x})$ and $\zeta(z)$ can be approximated by
\begin{subequations}\label{z_zeta_z_approximation}
\begin{align}
& z(\bar{\Omega},\vec{\bar{k}}_\perp,\bar{x}) \approx \frac{1}{|\bar{\Omega}|} [ 1  +  \bar{a} (\bar{k}_\perp^2 + \bar{x})]  ,\\
&  | \zeta(z(\bar{\Omega},\vec{\bar{k}}_\perp,\bar{x})) |^{3/2} \approx  \left| \zeta\left( \frac{1}{|\bar{\Omega}| } \right)\right|^{3/2}  \nonumber \\
& - \text{sign}(|\bar{\Omega}|-1) \frac{3}{2}\bar{a} \sqrt{|\bar{\Omega}^2 - 1|} (\bar{k}_\perp^2 + \bar{x}). \label{zeta_z_approximation}
\end{align}
\end{subequations}
Condition $||\tilde{\Omega}|-1| \gg \bar{a}$ ensures that the Taylor expansion (\ref{zeta_z_approximation}) is performed sufficiently far from the singularity $z=1$ of the derivatives of $|\zeta(z)|^{3/2}$.

If $||\bar{\Omega}| - 1| \sim \bar{a}$, Eq.~(\ref{z_zeta_z_approximation_Puiseux}) leads to $\zeta(z(\bar{\Omega},\vec{\bar{k}}_\perp,\bar{x})) \sim \bar{a}$, which means that the argument of the Airy function in Eq.~(\ref{F_tilde_Rindler_Hankel_2}) does not diverge. Specifically, Eq.~(\ref{F_tilde_Rindler_Hankel_2}) can be approximated by
\begin{align}\label{F_tilde_Rindler_Hankel_2_nonrelativistic}
& \bar{\tilde{F}} (\bar{\Omega},\vec{\bar{k}}_\perp, \bar{x})  \approx \frac{1}{2 \pi} \text{Ai} \left( \bar{k}_\perp^2 + \bar{x} - \frac{|\bar{\Omega}|  -1}{\bar{a}} \right),
\end{align}
which has already been proved for non-relativistic modes in the Rindler frame \cite{falcone2022non}.

If $||\bar{\Omega}| - 1| \gg \bar{a}$, divergences in the argument of the Airy function of Eq.~(\ref{F_tilde_Rindler_Hankel_2}) appear. Specifically, if $|\bar{\Omega}| - 1 \ll - \bar{a}$, then $|\bar{\Omega}| < 1 $ and $| \zeta(1/|\bar{\Omega}|) | \gg \bar{a}$. This means that the argument of the Airy function diverges at $+\infty$, leading to
\begin{equation}
\text{Ai}(\xi) \sim \frac{1}{\sqrt[4]{\xi}} \exp \left( -\frac{2}{3} \xi^{3/2} \right)
\end{equation}
and, hence, leading to an exponentially vanishing $\bar{\tilde{F}} (\bar{\Omega},\vec{\bar{k}}_\perp, \bar{x})$. Conversely, if $|\bar{\Omega}| - 1 \gg \bar{a}$, the argument of $\text{Ai}(\xi)$ diverges at $\xi \rightarrow -\infty$, leading to a rapidly oscillating Airy function. Indeed, modulus and phase of $\text{Ai}(\xi)$ have the following asymptotic leading terms
\begin{equation}\label{Airy_asymptotic_minus_infinity}
\text{Ai}(\xi) \approx \frac{1}{\sqrt{\pi}\sqrt[4]{-\xi}} \sin \left( \frac{\pi}{4} + \frac{2}{3}(-\xi)^{3/2} \right).
\end{equation}
Because of this rapidly oscillating behavior, the integral of Eq.~(\ref{chi_bar}) vanishes.

To explicitly shows that Eq.~(\ref{chi_bar}) is vanishing in the regime of $|\bar{\Omega}| - 1 \gg \bar{a}$, we use Eqs.~(\ref{F_tilde_Rindler_Hankel_2}) and (\ref{Airy_asymptotic_minus_infinity}) in Eq.~(\ref{chi_bar}):
\begin{align} \label{chi_bar_approximation_2}
& \bar{\chi}(\vec{\bar{k}}, \bar{\Omega}, \delta \bar{x}) \approx \frac{\bar{\Omega} + 1}{2 \pi \sqrt{|\bar{\Omega}| \delta \bar{x}}}  \int_{- \delta \bar{x}}^{\delta \bar{x}} d \bar{x}  \nonumber \\
& \times \sqrt[4]{\frac{2 \bar{a}}{(1 + 2 \bar{a} \bar{k}^2) [1-z^2(\bar{\Omega},\vec{\bar{k}}_\perp,\bar{x})]}}  \nonumber \\
& \times   e^{ -i \bar{k}_1 \bar{x} }  \sin \left( \frac{\pi}{4} + \frac{\sqrt{2}\bar{\Omega}}{3}  \left[ \frac{\zeta(z(\bar{\Omega},\vec{\bar{k}}_\perp,\bar{x}))}{\bar{a}} \right]^{3/2} \right),
\end{align}
which can be furthermore approximated by
\begin{align} \label{chi_bar_approximation_3}
& \bar{\chi}(\vec{\bar{k}}, \bar{\Omega}, \delta \bar{x}) \approx \frac{\bar{\Omega} + 1}{2 \pi \sqrt{\delta \bar{x}}}  \sqrt[4]{\frac{2 \bar{a}}{|\bar{\Omega}|^2 -1}}  \int_{- \delta \bar{x}}^{\delta \bar{x}} d \bar{x}   e^{ -i \bar{k}_1 \bar{x} }    \nonumber \\
& \times \sin \left( \frac{\pi}{4} +\frac{\sqrt{2}\bar{\Omega}}{3}  \left[ \frac{\zeta(z(\bar{\Omega},\vec{\bar{k}}_\perp,\bar{x}))}{\bar{a}} \right]^{3/2} \right).
\end{align}
By working in the regime (\ref{constraints_adimentional}), one can use Eq.~(\ref{zeta_z_approximation}) in order to see Eq.~(\ref{chi_bar_approximation_3}) having the following form
\begin{align} \label{chi_bar_approximation_4}
\bar{\chi}(\vec{\bar{k}}, \bar{\Omega}, \delta \bar{x}) \approx & \frac{\bar{\Omega} + 1}{2 \pi \sqrt{ \delta \bar{x}}}  \sqrt[4]{\frac{2 \bar{a}}{|\bar{\Omega}|^2-1}}  \int_{- \delta \bar{x}}^{\delta \bar{x}} d \bar{x}   e^{ -i \bar{k}_1 \bar{x} }   \nonumber \\
& \times  \sin ( -\kappa(\bar{\Omega}) \bar{x} + \varphi(\vec{\bar{k}}_\perp,\bar{\Omega}) ),
\end{align}
with
\begin{subequations} 
\begin{align}
\kappa(\bar{\Omega}) = & \frac{\bar{\Omega}}{\sqrt{2}}   \sqrt{\frac{\bar{\Omega}^2 - 1}{\bar{a}}},\label{kappa} \\
\varphi(\vec{\bar{k}}_\perp,\bar{\Omega}) = & \frac{\pi}{4} + \frac{\bar{\Omega}}{\sqrt{2}}   \left\lbrace \frac{2}{3} \left[ \frac{1}{\bar{a}} \zeta\left( \frac{1}{|\bar{\Omega}| } \right)\right]^{3/2}  \right.  \nonumber \\
& \left. -   \sqrt{\frac{\bar{\Omega}^2 - 1}{\bar{a}}} \bar{k}_\perp^2 \right\rbrace.
\end{align}
\end{subequations}

One can finally see that $\bar{\chi}(\vec{\bar{k}}, \bar{\Omega}, \delta \bar{x})$ in Eq.~(\ref{chi_bar_approximation_4}) is vanishing because of an infinitely rapidly oscillating Airy function integrated over a finitely oscillating function. Indeed, the two frequencies are respectively $\kappa(\bar{\Omega})$ and $\bar{k}_1$. While $\bar{k}_1$ is finite ($\bar{k}_1 \lesssim 1$), $\kappa(\bar{\Omega})$ diverges when $|\bar{\Omega}| - 1 \gg \bar{a}$.

We proved that when $||\bar{\Omega}| - 1| \gg \bar{a}$, $\bar{\chi}(\vec{\bar{k}}, \bar{\Omega}, \delta \bar{x})$ vanishes. One can consider the two following remaining cases: $|\bar{\Omega} - 1| \lesssim \bar{a}$ and $|-\bar{\Omega} - 1| \lesssim \bar{a}$. The case $|-\bar{\Omega} - 1| \lesssim \bar{a}$ has to be excluded because of the $\bar{\Omega} + 1$ factor appearing in Eq.~(\ref{chi_bar}), which makes $\bar{\chi}(\vec{\bar{k}}, \bar{\Omega}, \delta \bar{x})$ vanishing when $\bar{\Omega} \approx -1$. The only non-vanishing case is $|\bar{\Omega} - 1| \lesssim \bar{a}$. This concludes our proof on Eq.~(\ref{non_relativistic_Theta_1_adimensional}).

\section{} \label{appendix_2}

We prove Eq.~(\ref{alpha_tilde_approx_3}). Such proof follows from considering any function $\varphi(\xi)$ in $\xi>0$ and the following integral
\begin{widetext}
\begin{align} \label{Kontorovich_Lebedev_0}
&\int_0^\infty dx \int_0^\infty d\Theta_1  \frac{2 \Theta_1}{\hbar c^2 a x}  \tilde{F}(\vec{\Theta},X_\text{R}(x)) \tilde{F}(\vec{\Theta},X) \varphi\left( \sqrt{c^2 \Theta_\perp^2 + \left(\frac{mc^2}{\hbar}\right)^2} \frac{x}{c} \right)\nonumber \\
 = & \frac{1}{2\pi^4 (c a)^2} \int_0^\infty dx \int_0^\infty d\Theta_1  \frac{\Theta_1}{x} \sinh \left( \frac{\beta \Theta_1}{2} \right)   K_{i \Theta_1 / (c a)} \left( \sqrt{c^2 \Theta_\perp^2 + \left(\frac{mc^2}{\hbar}\right)^2} \frac{x}{c} \right)  \nonumber \\
&  \times K_{i \Theta_1 / (c a)} \left( \sqrt{c^2 \Theta_\perp^2 + \left(\frac{mc^2}{\hbar}\right)^2} \frac{e^{aX}}{c a} \right)   \varphi\left( \sqrt{c^2 \Theta_\perp^2 + \left(\frac{mc^2}{\hbar}\right)^2} \frac{x}{c} \right).
\end{align}
\end{widetext}

By using the coordinate transformation
\begin{align}
& \zeta = \frac{\Theta_1}{ca}, & \xi = \sqrt{c^2 \Theta_\perp^2 + \left(\frac{mc^2}{\hbar}\right)^2} \frac{x}{c},
\end{align}
Eq.~(\ref{Kontorovich_Lebedev_0}) reads
\begin{align} \label{Kontorovich_Lebedev_1}
&\int_0^\infty dx \int_0^\infty d\Theta_1  \frac{2 \Theta_1}{\hbar c^2 a x}  \tilde{F}(\vec{\Theta},X_\text{R}(x)) \tilde{F}(\vec{\Theta},X)  \nonumber \\
& \times \varphi\left( \sqrt{c^2 \Theta_\perp^2 + \left(\frac{mc^2}{\hbar}\right)^2} \frac{x}{c} \right) \nonumber \\
=  &  \frac{1}{2 \pi^4} \int_0^\infty d\xi \int_0^\infty d\zeta \frac{\zeta}{ \xi} \sinh ( \pi \zeta )  K_{i \zeta} (\xi)  \nonumber \\
& \times K_{i \zeta} \left( \sqrt{c^2 \Theta_\perp^2 + \left(\frac{mc^2}{\hbar}\right)^2} \frac{e^{aX}}{c a} \right) \varphi (\xi).
\end{align}
Equation (\ref{Kontorovich_Lebedev_1}) can also be written in the following way
\begin{align}\label{Kontorovich_Lebedev_2}
&\int_0^\infty dx \int_0^\infty d\Theta_1  \frac{2 \Theta_1}{\hbar c^2 a x}  \tilde{F}(\vec{\Theta},X_\text{R}(x))  \tilde{F}(\vec{\Theta},X) \nonumber \\
& \times \varphi\left( \sqrt{c^2 \Theta_\perp^2 + \left(\frac{mc^2}{\hbar}\right)^2} \frac{x}{c} \right) \nonumber \\
= & \frac{1}{4 \pi^2}  \mathcal{K}^{-1} [\mathcal{K} [\varphi]] \left( \sqrt{c^2 \Theta_\perp^2 + \left(\frac{mc^2}{\hbar}\right)^2} \frac{e^{aX}}{c a} \right),
\end{align}
where
\begin{equation}
\mathcal{K} [\varphi] (\zeta) = \frac{2 \zeta}{\pi^2} \sinh(\pi \zeta) \int_0^\infty \frac{d\xi}{\xi}  K_{i \zeta}(\xi)  \varphi(\xi)
\end{equation}
is the Kontorovich–Lebedev transform and
\begin{equation}
\mathcal{K}^{-1} [\varphi] (\xi) = \int_0^\infty d\zeta  K_{i \zeta}(\xi)  \varphi(\zeta)
\end{equation}
its inverse. Since $\mathcal{K}^{-1}$ is the inverse of $\mathcal{K}$, Eq.~(\ref{Kontorovich_Lebedev_2}) reads
\begin{align}\label{Kontorovich_Lebedev_3}
&\int_0^\infty dx \int_0^\infty d\Theta_1  \frac{2 \Theta_1}{\hbar c^2 a x}  \tilde{F}(\vec{\Theta},X_\text{R}(x)) \tilde{F}(\vec{\Theta},X)  \nonumber \\
& \times \varphi\left( \sqrt{c^2 \Theta_\perp^2 + \left(\frac{mc^2}{\hbar}\right)^2} \frac{x}{c} \right) \nonumber \\
= & \frac{1}{4 \pi^2}  \varphi \left( \sqrt{c^2 \Theta_\perp^2 + \left(\frac{mc^2}{\hbar}\right)^2} \frac{e^{aX}}{c a} \right).
\end{align}

Since Eq.~(\ref{Kontorovich_Lebedev_3}) holds for any $\varphi$, we have proven Eq.~(\ref{alpha_tilde_approx_3}).

\bibliography{bibliography}

\end{document}